\documentclass[apj]{aeb_emulateapj_2010}

\usepackage{amsmath}
\usepackage{amssymb}
\usepackage{mathrsfs}

\usepackage[usenames,dvips]{color}

\usepackage{natbib}
\def\Po{P_{\rm osc}}
\def\Eo{E_{\rm osc}}
\def\zc{\zeta_{\rm crit}}
\def\Eb{E_{\rm b}}
\def\Et{E_{\rm t}}

\def\Sc{\Sigma_{\rm crit}}

\newcommand\bmath[1] {\mbox{\boldmath$\rm #1$}}

\def\e{{\rm e}}

\def\yr{{\rm yr}} 

\begin{document}

\title{The evolution of cloud cores and the  formation of stars}

\author{
Avery E.~Broderick\altaffilmark{1} \& 
Eric Keto\altaffilmark{2}
}
\altaffiltext{1}{Canadian Institute for Theoretical Astrophysics, 60 St.~George St., Toronto, ON M5S 3H8, Canada; aeb@cita.utoronto.ca}
\altaffiltext{2}{Smithsonian Observatory, 60 Garden St., Cambridge, MA 02138, USA; keto@cfa.harvard.edu}

\shorttitle{Oscillation-Delayed Collapse and Star Formation}
\shortauthors{Broderick \& Keto}

\begin{abstract}
For a number of starless cores, self-absorbed molecular line and
column density observations have implied the presence of
large-amplitude oscillations.  We examine the consequences
of these oscillations on the evolution of the cores and
the interpretation of their observations. We find that 
the pulsation energy helps support the cores and that
the dissipation of this energy can lead toward instability
and star formation. 
In this picture, the core lifetimes are limited by the pulsation decay
timescales, dominated by non-linear mode-mode coupling, and on the
order of $\simeq{\rm few}\times10^5$--$10^6\,\yr$.  Notably, this is
similar to what is required to explain the relatively low rate 
of conversion of cores into stars.
For cores with large-amplitude oscillations, dust continuum
observations may appear asymmetric or irregular. As a consequence,
some of the cores that would be classified as supercritical may be
dynamically stable when oscillations are taken into account.
Thus, our investigation motivates a simple hydrodynamic picture,
capable of reproducing many of the features of the progenitors of
stars without the inclusion of additional physical processes, such as
large-scale magnetic fields.
\end{abstract}

\keywords{hydrodynamics--ISM: clouds -- ISM: globules -- stars: formation}

\maketitle

\section{Introduction}\label{sec:I}

The starless cores owe their reputation as the sites of future star formation to the similarities
they share with neighboring cores already in the pains of stellar birth.
Small (tenths of pc), dense  ($n_{\rm H_2} \sim 10^3$ to $10^6$ cm$^{-3}$) dark clouds
of a few solar masses, the starless cores contain no infrared sources above 
the sensitivity level of the IRAS satellite (about 0.1 L$_\odot$ at the distance of Taurus).
Mixed in with a population of cores, embedded infrared protostars, and young  T-Tauri 
stars, the starless cores seem on the verge of issue. 
\citep{MyersLinkeBenson1983, MyersBenson1983, BensonMyers1989, Beichman1986,
DiFrancesco2007, Ward-Thompson2007,BerginTafalla2007}.

The starless cores are remarkable in another sense. They may be the
only structures in the interstellar medium in quasi-static equilibrium with
lifetimes longer than their sound-crossing times
\citep{Beichman1986,JessopWT2000,KirkWTAndre2005,Ward-Thompson2007}. Whereas
all larger-scale clouds are better described as transient structures 
within a turbulent cascade \citep{Field2008}, 
the starless cores show observed densities that
are consistently matched 
\citep{Ward-Thompson1994,Andre1996,Ward-Thompson1999,Bacmann2000,Evans2001,AlvesLadaLada2001,Tafalla2002,Kandori2005,KirkWTAndre2005}
by the truncated solutions of the isothermal Lane-Emden equation,
i.e., the Bonnor-Ebert (BE) spheres \citep{Bonnor1956}.  These are
self-gravitating spheres supported by thermal energy and bounded by an
external pressure.  The source of the external pressure appears to
depend upon the star forming region, though its necessity is well
documented.  Within the Pipe nebula, the core energetics are
consistent with a constant external pressure supplied by hotter, more
rareified gas \citep{Lada2008}.  In other regions, such as Lupus
\citep{Teixeira2005}, the cores appear to be surrounded by lower
density molecular gas with observed molecular line widths $\geq 1$
kms.  Here and in similar regions small-scale (smaller than the
observing beam) supersonic turbulence in the low-density
molecular gas around the cores may
supply the confining pressure.

While a continuous extension to the larger-scale ISM prompts the
question of whether the cores are individual dynamical entities, a
distinct boundary is evident in several observations.  First, maps of
dust absorption in the Taurus region \citep{Bacmann2000} often show
sharp edges defining the core boundaries.  Second, around at least one
core, B5\footnote{The core B5 contains a protostar but is otherwise
  similar to the starless cores. Since the effects of the protostar
  are not significant at the core boundary, the same demarcation would
  be expected in the truly starless cores.}, a boundary is defined by
a sharp transition (within less than an observing  beam width) between
subsonic and supersonic turbulence in the molecular gas
\citep{Pineda2010}.  Third, a boundary is a necessary feature in
models  of radiative cooling and UV heating
\citep{ Evans2001,ShirleyEvansRawlings2002,Zucconi2001,StamatellosWhitworth2003,Goncalves2004,KetoField2005}
that use the basic BE structure and accurately predict the observed
gas and dust temperatures
\citep{Ward-Thompson2002,Pagani2003,Pagani2004,Crapsi2007}, the
observed variation in the excitation of the CO molecule across the
starless cores \citep{Bergin2006,Schnee2007,KetoCaselli2008}, and the
observed brightness and profiles of molecular lines
\citep{KetoCaselli2010}.

The predictive success of models based on BE spheres (including a core
boundary) implies that the starless cores are indeed individual
dynamic entities in quasi-static equilibrium, balancing the forces of
their own self-gravity, internal energy, and the larger-scale ISM. In
some cases, this equilibrium may be unstable, leading the core to
contract, though their internal pressure is sufficiently high to
prevent free-fall.

The hydrodynamic equations that describe the equilibrium configuration
of both stars and the starless cores allow for perturbations
in the form of sonic waves that may persist for many crossing times.
Such oscillations are observed in the Sun, other main sequence stars,
and white dwarfs, and given the turbulent state of the larger-scale ISM,
should be expected in the starless cores as well.
Analytic \citep{Keto2006} and numerical
\citep{KetoField2005,Broderick2007} models of
oscillating cores easily reproduce the characteristic asymmetric shapes of 
molecular lines seen in the starless cores. 
While the spectral-line profiles of the 
low-order modes (the dipolar and breathing modes are
observational indistinguishable from simple contraction, expansion and
rotation) a few cores show simultaneous inward and outward motions
\citep{Aguti2007,Lada2003,Redman2006} that are consistent only with
oscillatory modes.  Oscillations produce density perturbations as
well, and the dipole and quadrupole modes naturally reproduce the
ellipsoidal morphologies \citep{Keto2006} characteristically seen in 
observations of the cores \citep{Bacmann2000,Ward-Thompson1999,KirkWTAndre2005}.

Observations suggest that these oscillations, rather than a minor
perturbation on quasi-static equilibrium, provide a substantial
contribution to the dynamical stability and therefore are important in
the evolution of the cores toward star formation.  First, a significant fraction of
starless cores have observed density profiles that can only be matched
by a BE sphere if the gas temperature in the model sphere is made
considerably higher than is known to be the case in the starless cores
\citep{Evans2001,KirkWTAndre2005,Kandori2005}.  Second, the sizes of
many of the starless cores imply that they should be unstable to
gravitational collapse \citep{Evans2001,Kandori2005}. However,
statistical analyses of the relative numbers of stars and cores in a
sample of different star-forming regions imply lifetimes that are
several times the relevant free-fall time
\citep{Beichman1986,JessopWT2000,Ward-Thompson2007}.  Reconciling
these observations require a source of energy within the cores in
addition to the thermal energy.  Internal oscillations or turbulence
are suggested by the observation that the spectral line widths in many
cores are larger than the expected thermal width
\citep{DickmanClemens1983,MyersBenson1983,Lada2008} implying internal
subsonic velocities on spatial scales unresolved in the observing
beam.

These observations combined with modelling suggest the following
evolution for the starless cores:  The cores are formed at the bottom
of the supersonic turbulent cascade that dominates the larger-scale
ISM \citep{Field2008}, producing cores with a spectrum of masses and
internal turbulent energies.  Those cores massive enough to be
gravitationally unstable immediately begin contraction to form a star,
while those with appropriate mass and internal energy become
self-supported and self-gravitating. These cores are decoupled from
the cascade in the sense that they exist as individual dynamical
entities independent of the larger-scale flows \citep{KetoField2005}.
The internal turbulence inherited from the parent flow, and present
when a core forms, is part of its initial dynamical stability but is
expected to dissipate over time.  Thought of as a spectrum of
individual modes of oscillation, the turbulence decays by non-linear
mode-mode coupling (collisions between modes) \citep{Broderick2007}
and by transmission through the core boundary into the larger-scale
ISM \citep{Broderick2008}.  The gradual loss of this internal
turbulent energy leads eventually (on time scales of several crossing
times) to gravitational instability and star formation.

In this paper we explore the latter half of this evolution.  We
develop a model for the internal turbulence of the cores and examine
its consequences for the evolution toward star formation and for the
interpretation of dust continuum observations.  Using the fact that
the turbulence can be fully described as a collection of oscillating
modes, we consider three models for the turbulent energy distribution,
a Kolmogorov spectrum, a flat spectrum, and a spectrum dominated by
long wave lengths.  In Section \ref{sec:AME} we compute the effective
pressure arising from this turbulence.  We find that the breathing
modes are destabilizing, but modes with shorter wave lengths generate
a supporting pressure.  In Section \ref{sec:MSEC}, we generate random
realizations of turbulent cores and find that a typical stable,
oscillating core would appear to be dynamically unstable, if the
observations were fit to a static BE model.
In Section \ref{sec:EoaMSSC}, we follow the evolution of the turbulent
spectrum as it decays. We find that the spectrum of the turbulence
evolves toward dominance by the longer wave lengths.  In conclusion,
we find that we can describe the evolution of the cores toward star
formation as a purely hydrodynamic phenomenon without the necessity of
support by large-scale magnetic fields (though a tangled magnetic
field may co-exist in equipartition with the turbulence without
substantially changing the findings).

\section{The Stability of Pulsating Cores}\label{sec:AME}
\begin{figure}
\begin{center}
\includegraphics[width=\columnwidth]{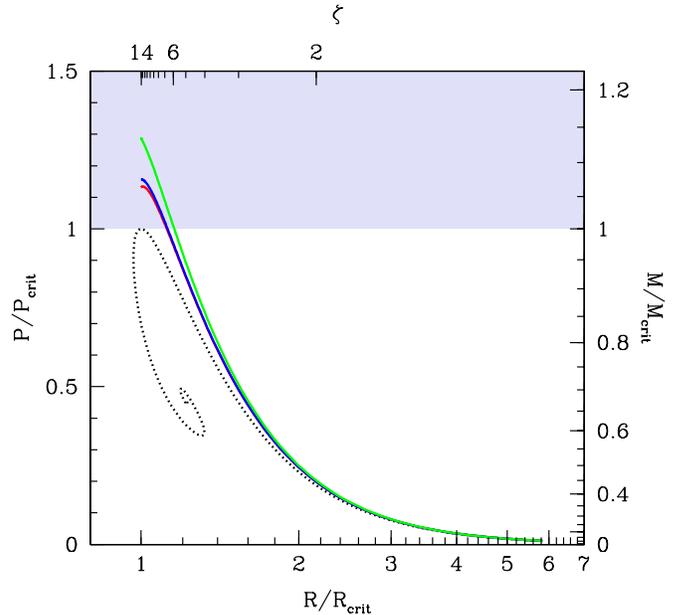}
\end{center}
\caption{Surface pressure as a function of radius for a pulsating
  fixed-mass isothermal gas sphere.  For reference, the static
  configuration is shown by the dotted line, and becomes unstable at a
  center/surface density ratio, $\zeta$, of $14.04$.  The red, blue
  and green lines shown the surface pressure resulting from
  oscillations containing 30\% of the unperturbed configurations
  gravitational binding energy for the Kolmogorov (Equation
  \ref{eq:KES}), Flat (Equation \ref{eq:FES}) and
  fundamental-Kolmogorov (Equation \ref{eq:fKES}) energy spectra,
  respectively.  Along the top axis $\zeta$ is shown until the
  critical value is reached (above which the mapping from $\zeta$ to
  $R$ is no longer single valued, as evidenced by the spiral structure
  in $P$ vs. $R$ for $\zeta>\zc$).  Along the right axis the
  corresponding mass at fixed surface pressure is shown.  Finally,
  within the shaded region at the top no stable static configurations
  exist.
}\label{fig:PvR}
\end{figure}

In this section we compute the dynamical consequences of large-scale
oscillations for the underlying starless core.  We find it is
possible for pulsations both to support the core  and to
initiate collapse, the distinction lying in the relative importance
of the low and higher-order oscillation modes.  It is possible
to do this with a macroscopic description 
similar to the stability analysis of a BE sphere  \citep{Bonnor1956},
requiring only the pulsation frequencies (which depend upon the core structure)
and the distribution of energy amongst them.  Here we briefly review
how stability may be determined by the macroscopic core properties,
make explicit the notion of ``slow'' compression, use this definition to
compute the response of the pulsations, and finally assess the
consequences for a number of different mode compliments.

\subsection{The Stability of Stationary Cores}
The starless cores in the BE model are held together by a
combination of their own self-gravity and the external pressure, and
these two inward forces are resisted by the outward thermal pressure.
The dotted line in Figure \ref{fig:PvR} shows the equilibrium configurations
as a function of the external pressure,$P$, and the core size, $R$.
At large radii, the portion of the curve to the right of $R/R_{crit} = 1$,
the gravitational force at the boundary is weak compared to
the confining pressure. If we compress a core on this portion of the
curve, we find that the internal  pressure increases with the 
increasing density, and a larger external pressure is required to
confine the core, $dP/dR<0$. If the compression is removed,
the core will re-exand to its original equilibrium. 
However, if the radius becomes sufficiently small, $R /R_{crit} < 1$,
gravitational force dominate at the boundary,  and a smaller external
pressure would be required to maintain equilibrium as the core is
compressed, i.e., $dP/dR\ge0$. In this case, the core can not resist
further compression and is therefore unstable to contraction.
For static isothermal gas spheres, this is greatly simplified by the
fact that the equilibrium configurations form a one-parameter family,
which may be chosen such that the relevant parameter is the ratio of
the surface and central densities (or, equivalently, pressures),
$\zeta$ \citep{Bonnor1956}.  In this case the maximum ratio is
$\zeta_{crit} = 14.04$.

The thought experiment we describe above to define the stability of cores
employs a sequence of cores at fixed mass and varying surface
pressure.  However, it is perhaps more physically appropriate to hold
the surface pressure fixed, determined by the properties of the
surrounding molecular gas, and vary the core mass instead.  That is,
at a given surface pressure, core radius and core temperature identify
the unique, stable isothermal core mass, $M(R)$ (if an equilibrium
configuration exists at all).  The mass and pressure are related by
the scaling relations that connect gas spheres with identical
$\zeta$'s.  Explicitly, from the equations of hydrostatic structure,
at a given $\zeta$ we have $M\propto c_s^3/P^{1/2}$, and thus fixing
$c_s$ (i.e., fixing the core temperature) gives $M\propto P^{-1/2}$.
The $P(R)$ shown in Figure \ref{fig:PvR} are computed at fixed mass,
i.e., $M={\rm const}$.  Conversely, if we choose $P={\rm const}$, we
have
\begin{equation}
\frac{M(R)}{M} = \left[\frac{P(R)}{P}\right]^{1/2}
\end{equation}
Thus, the equilibrium core mass $M(R)$ at fixed surface pressure and core
temperature is identical, up to a non-linear rescaling, to $P(R)$.  In
Figure 1 the right-hand vertical axis is labeled with $M(R)$ instead
of $P(R)$.  Most importantly, $P_{\rm crit}$ corresponds to a critical
mass, $M_{\rm crit}$, above which no stable configuration exists.

\subsection{Response of Pulsations to Slow Compression} \label{sec:RoPtSC}
The situation becomes more complicated in the presence of
energetically significant pulsations.  This is because we must not
only adjust the underlying gas configuration, but also the oscillation
amplitudes.  In principle, a rapid compression can excite oscillations
itself, and we must carefully specify the dynamical nature
of the experiment we use to define stability.  More generally, we must
carefully identify how ``equilibrium'' configurations are defined
along the entire $P(R)$ curve.  In the static case a similar
consideration is already made when we assume that the core mass and
temperature are fixed; both are natural physical conditions for the
starless cores. Since the core has a boundary, its mass is
constant\footnote{We allow that real cores may occasionaly gain mass
  from interaction with the surrounding ISM, but this is not a
  necessary process for this discussion.}. Radiation provides an
efficient thermal coupling to the surrounding medium keeping the core
temperature unvarying on a time scale much shorter than the dynamical
time scales of the cores \citep{KetoCaselli2010}.

Although radiative equilibrium violates the adiabatic condition in the strict
thermodynamic sense, if the compression is assumed to occur slowly
relative to the oscillation periods, we may still treat it as
adiabatic for the purposes of evolving the oscillation mode
amplitudes.  In this case, what remains fixed are the adiabatic
invariants\footnote{The notion of an adiabatic 
  invariant arises naturally within the context of the Hamilton-Jacobi
  formulation of classical mechanics.  A full discussion of adiabatic
  invariants and their relationship to the action-angle variables of
  the Hamilton-Jacobi theory may be found in \S11-7 of
  \citet{Goldstein1980} or \S50 and \citet{LandauLifshitz1976}.  Of
  particular relevance are the adiabatic invariants of the simple
  harmonic oscillator, which is presented as an example in both
  references.} of the oscillations.  Computing these is simplified by the 
fact that we may treat the pulsation modes as a collection of
uncoupled, one-dimensional harmonic oscillators, with the mode amplitudes
defined by the radial and angular quantum numbers $n$, $l$ and
$m$ (with their standard definitions), $A_{nlm}$, governed by
\begin{equation}
\frac{d^2 A_{nlm}}{dt^2} = -\omega_{nlm}^2 A_{nlm}\,,
\end{equation}
for a mode dependent angular frequency, $\omega_{nlm}$
\citep{Ging-Mona:80,Rath-Brod-Blan:03}.  For this the adiabatic
invariant is well known to be $E_{nlm}/\omega_{nlm}$, where
$E_{nlm}=M_{nlm}\omega_{nlm}^2A_{nlm}^2/2$ is the mode energy (in
which $M_{nlm}$ is the mode ``mass'' and computed from the mode
structure directly).  Thus, we may compute the evolution of the energy
in a particular mode during an adiabatic compression, given
$\omega_{nlm}(R)$.  Specifically, from the adiabatic invariant we
obtain 
\begin{equation}
\frac{dE_{nlm}}{dR} = \frac{E_{nlm}}{\omega_{nlm}} \frac{d\omega_{nlm}}{dR}\,.
\end{equation}
However, the presence of an additional component to the core's total
energy associated with the oscillations, $\Eo=\sum_{nlm}E_{nlm}$, also
implies the existance of an additional component to the surface
pressure.  That is, generally, under adiabatic compressions the
oscillations will produce a dynamically generated surface pressure
\begin{equation}
\Po
\equiv
-\frac{d\Eo}{dV}
=
-\frac{d}{dV}\sum_{nlm} E_{nlm}
=
-\frac{1}{4\pi R^2} \sum_{nlm} \frac{E_{nlm}}{\omega_{nlm}} \frac{d\omega_{nlm}}{dR}\,.
\label{eq:P}
\end{equation}
Thus, the pulsation contribution to the surface pressure is dependent
upon $d\omega_{nlm}/dR$ and the oscillation energy spectrum (which we
shall discuss in more detail below).

In the dicussion thus far, we have not placed any requirements upon
the underlying static configuration which define the structure of the
pulsations.  Therefore, we can gain some insight into what we
might expect from a simple, one-dimensional example, a uniform
temperature and density cubical cavity of side-length $2R$.  In
this case the wavelength of each oscillation mode, the standing waves
of the cavity, is trivially proportional to $R$ and corresponds to a
frequency proportional to $R^{-1}$.  Thus we find
$d\omega_{nlm}/dR = -R^{-2}$ and $\Po\propto R^{-4}\propto V^{-4/3}$,
i.e., the phonons constitute a $\Gamma=4/3$ gas.  Of course, this may
be computed directly from the dispersion relation $\omega = c_s k$
where $c_s$ is fixed.  Nevertheless, we may expect short-wavelength
pulsations to provide an additional component with a nearly identical
equation of state to that of small-scale, tangled magnetic fields.

\begin{figure}
\begin{center}
\includegraphics[width=\columnwidth]{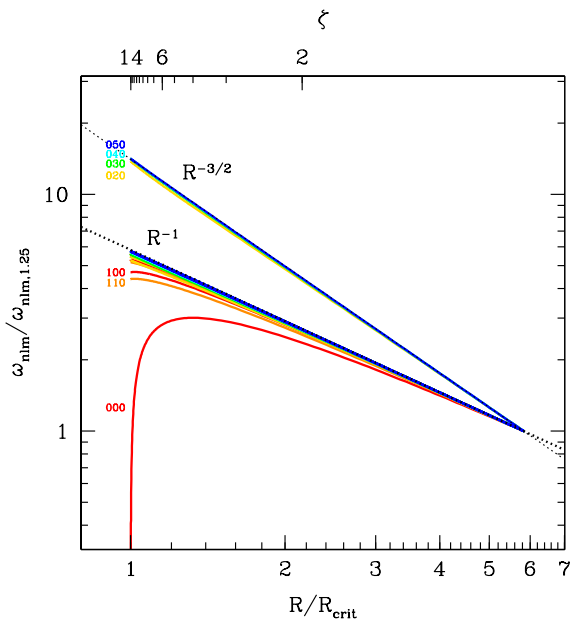}
\end{center}
\caption{Normalized pulsation frequencies for a number of low-harmonic
  oscillations of the isothermal gas sphere as a function of sphere
  radius, $R$.  The thick and thin black dotted line shows the
  $R^{-1}$ and $R^{-3/2}$, respectively.  While $n=0$--$5$ and
  $l=0$--$5$ are shown (the frequencies are degenerate in $m$), only
  those that may be easily distinguished from the $R^{-1}$ asymptotic behavior
  are labeled.  Frequencies with $l=0$, $1$, $2$, $3$, $4$ and $5$ are
  colored red, orange, yellow, green, cyan and blue, respectively.
}\label{fig:omegas}
\end{figure}

Generally, $\omega_{nlm}(R)$ will depend upon the structure of the
equilibrium core, in particular upon the relative sizes of the
core and the wave length of the oscillations.
For the oscillation modes of isothermal cores, we may construct the
mode structures and compute the mode frequencies directly, shown in
Figure \ref{fig:omegas} for a variety of radii
\citep[or, equivalently, $\zeta$; see, e.g.,][]{Rath-Brod-Blan:03,Keto2006}.
High-order modes, i.e., modes with wavelengths much shorter than the
core radius, don't probe the gas density gradients, and we recover
$\omega_{nlm}(R)\propto R^{-1}$.  In contrast, low-order modes do show
substantial departures from the $R^{-1}$ dependence, becoming most
significant for the fundamental mode ($n=0$).  For high meridional
wave number ($l>1$) the fundamental modes have
$\omega_{0lm}\propto R^{-3/2}$, i.e., the pulsation frequencies are
roughly proportional to the Keplerian frequency at the core surface,
$\omega_K\equiv(GM/R)^{3/2}$, corresponding to an adiabatic index
of $3/2$.  This is not surprising as these modes probe the radial
structure of the isothermal core. 

The only oscillation mode substantially deviating from either the
$R^{-1}$ or $R^{-3/2}$ behaviors is the breathing mode ($n=0$,
$l=0$ and $m=0$), responsible for the onset of instability at
$\zeta=\zc$, at which which $\omega_{000}$ vanishes.  Near $\zc$,
$d\omega_{000}/dR$ is positive and diverges, yielding a formally
infinite {\em negative} surface pressure.

\subsection{The Oscillation Energy Spectrum} \label{sec:TOES}

Once an energy spectrum, $E_{nlm}$, has been specified, we may produce a
revised $P(V)$ which includes the contributions from the oscillations.
This is now an explicit function of the mode amplitudes as well as $\zc$,
and thus no longer has a unique shape.  In setting the $E_{nlm}$
we will have intrinsically made some statement about the mode excitation
and evolution.  Indeed, as we shall see in Section \ref{sec:EoaMSSC},
the evolution tends to move power from small wavelengths to large
wavelengths.  Despite this, we may bound the likely possibilities, and
already a number of important qualitative statements can be made about
the dynamical importance of pulsations.

\subsubsection{Large-Amplitude Monopole Oscillations}
The breathing mode is unique in that, in principle, it can produce a
negative contribution to the surface pressure, diverging at
$\zeta=\zc$.  The breathing mode corresponds to a simple expansion
and contraction of the core, similar to changing $\zeta$ in the
family of equilibrium configurations. On the inward pulse, the
contraction may result in $\zeta > \zeta_{crit}$ and collapse of
the core. This can potentially destabilize otherwise stable
isothermal cores, inducing collapse at a lower external pressure (or
equivalently, smaller core mass).  However, the mathematical analysis of the
consequences of energetic breathing modes specifically, and monopole
pulsations more generally, are not clear for a number of reasons.

Foremost among these is the failure of the linear description of
the monopole oscillations.  That this must happen near $\zeta=\zc$ for
the breathing mode is suggested by the fact that $\omega_{000}$ vanishes
at this point, and thus
$A_{000} = \omega_{000}^{-1}\sqrt{2 E_{000}/M_{000}}$ diverges for finite
mode energies ($M_{000}$ is not pathological at $\zc$).  More
importantly, the density perturbations associated with the breathing
mode diverge, doing so most strongly at the core center.  Restricting
our attention to pulsation amplitudes, and thus energies, for which
the central density perturbation is less than the unperturbed central
density (a somewhat permissive definition of
``perturbation'') results in fractional changes in the supportable
external pressure of less than 0.5\% at all $\zeta$, and half that at
most $\zeta$.  That is, when we consider mode amplitudes that are
indeed perturbative, the dynamical effect of the breathing mode is
negligible.  A similar constraint limits the dynamical importance of
the higher-order monopole pulsations.  Unlike the breathing mode, these
generally increase the allowed external pressure, as initially
anticipated.  Nevertheless, as with the breathing mode, limiting their
amplitudes to the perturbative regime results in a small correction
to the stability of isothermal cores.

What happens in the non-linear regime is not {\em a priori} clear, and
beyond the scope of this paper.  However, we might expect that 
cores with oscillations dominated by non-linear monopole pulsations
are short-lived.  If non-linear monopole oscillations can indeed be
excited in a subset of starless cores, and these do induce instability
in a fashion similar to that anticipated by the linear analysis of the
breathing mode given above, then these cores will either collapse or
disperse on a single dynamical time.  These cannot be the apparently
long-lived cores that are observed.  That is, the observed cores are
necessarily those without large-scale breathing modes.

Alternatively, large-amplitude breathing modes may simply be uncommon
in starless cores.  If the pulsations are driven by the turbulence
from the parent molecular cloud, it may be difficult to excite monopole
oscillations.  The intra-core motions are typically subsonic, and at
the core scale, the turbulence is therefore nearly incompressible.
However, the monopole modes generally, and the breathing mode
specifically, are primarily compressive oscillations.  Thus, the
molecular cloud turbulence can only weakly couple to these
oscillations, either through a small failure of the incompressible
condition or through non-linear mode coupling.

For these reasons we will not consider monopole pulsations further,
restricting our attention henceforth to oscillations with $l\ge1$.

\subsubsection{Non-radial Oscillations}
Higher-order pulsations are nearly always stabilizing, i.e., they
increase the allowed surface pressure at which stable isothermal cores exist,
or alternatively increase the mass that a core can support.  The
exception being the $n=1$, $l=1$ oscillation which produces a small
negative pressure near $\zc$\footnote{The fundamental dipole mode,
  $n=0$ \& $l=1$, does not exist as a consequence of momentum
  conservation.}.  From Equation (\ref{eq:P}) and the asymptotic
behaviors of the $\omega_{nlm}$ for the non-radial oscillations, it is
straightforward to estimate the contribution to the surface pressure
from the oscillations:
\begin{equation}
\Po
\sim
\frac{\Eo}{V}
=
\left|\frac{\Eo}{\Eb}\right| \frac{\Eb}{V}
\sim
\left|\frac{\Eo}{\Eb}\right| P_0\,,
\end{equation}
where $P_0$ is the surface pressure of the unperturbed configuration.
The precise constant of proportionality depends upon $\zeta$ and 
weakly on the energy
spectrum.

Because for $l>0$ the form of $\omega_{nlm}(R)$ is well approximated
by either $R^{-1}$ ($n>1$) or $R^{-3/2}$ ($n=0$), in practice it is
not necessary to fully specify the $E_{nlm}$.  Rather, it is sufficient
to define $\Eo^0=\sum_{n=0,l>0,m} E_{nlm}$ and
$\Eo^{>0}\equiv\sum_{n\ge1,l>0,m} E_{nlm}$, corresponding to the
energy in fundamental oscillations and the p-modes, respectively.  The
resulting excess pressure due to the pulsations is then
\begin{equation}
\Po
\sim
\frac{\Eo^0}{2V} + \frac{\Eo^{>0}}{3V}\,,
\end{equation}
where $V=4\pi R^3/3$ and the difference between the denominators
arises from the different asymptotic behaviors of the
$\omega_{nlm}(R)$.  Note that this is confined to the somewhat narrow
range from $\Eo/3V\le\Po\le\Eo/2V$.  While we will indeed specify the
full energy spectrum, the distinction between different energy spectra
will primarily be in the relative importance of $\Eo^0$ and $\Eo^{>0}$.

The particular form of the energy spectrum reflects both the mechanism
by which the pulsations are excited and their subsequent evolution
within the core.  Here we consider three different forms for the
turbulent energy spectrum, seeking to bracket the possible
consequences.  
A natural
place to begin is to assume a Kolmogorov spectrum for the
oscillations, i.e.,
\begin{equation}
E_{nlm}^K\propto
\left\{\begin{array}{ll}
\displaystyle k_{nlm}^{-11/3} & \quad {\rm if~}l>0\\
0 & \quad{\rm otherwise}
\end{array}
\right.
\label{eq:KES}
\end{equation}
where
\begin{equation}
k_{nlm}=\frac{\pi}{R}\sqrt{1+n^2+l(l+1)}\,,
\end{equation}
so that $E_k = 4\pi k_{nlm}^2 E_{nlm}^K \propto k^{-5/3}$.  Despite
the heavy weight upon long-wavelengths, this is dominated by
$n>0$ modes, primarily due to the dipole and quadrupole oscillations,
such that we have $\Eo^{>0}/\Eo^0\simeq3.5$.  For concreteness, in
Figure \ref{fig:PvR}  the resulting surface pressure is shown by the
red line for $\Eo=0.3\Eb$, though this scales linearly with $\Eo/\Eb$.

\begin{figure*}
\begin{center}
\includegraphics[width=0.3284\textwidth]{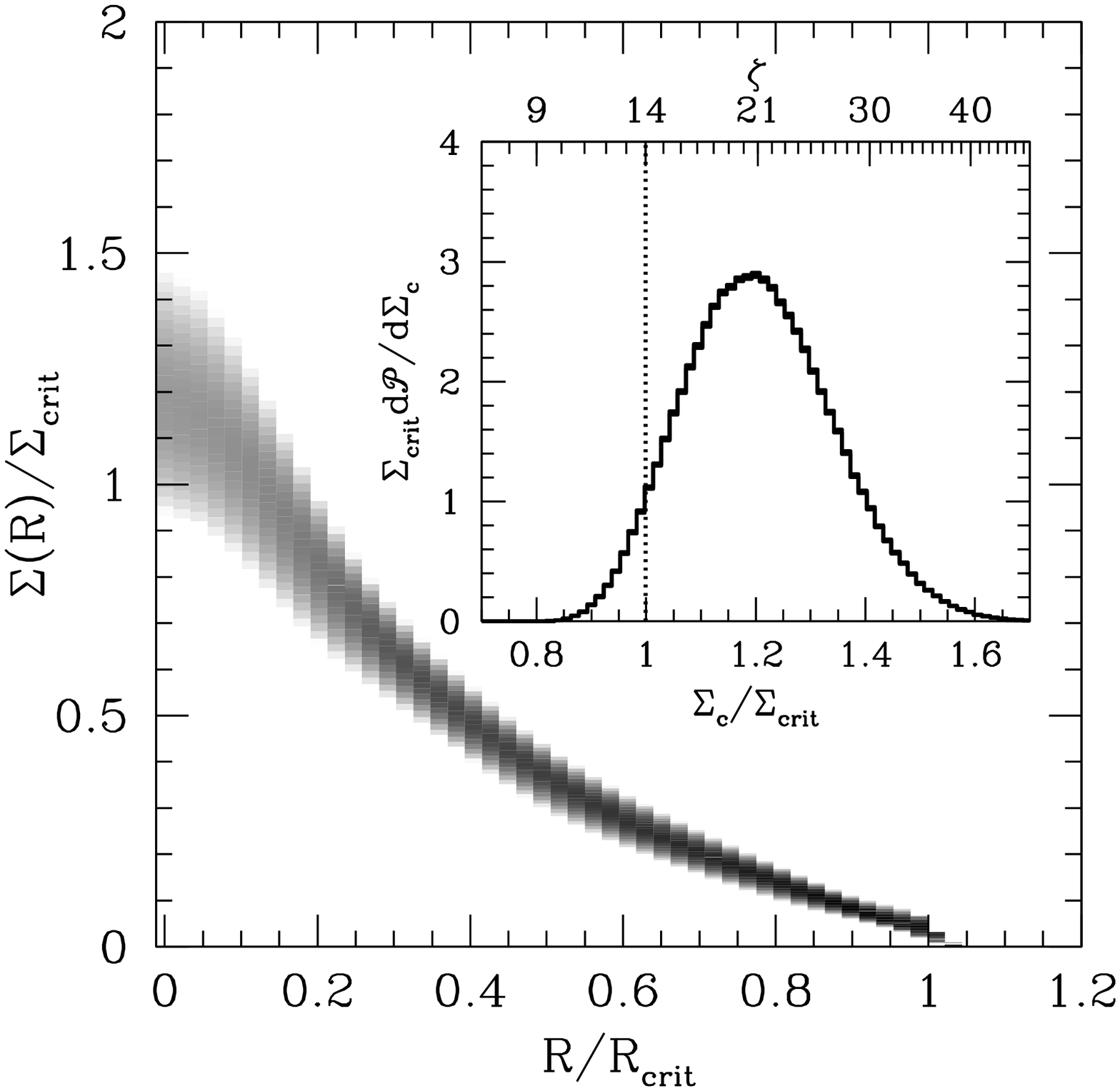}
\includegraphics[width=0.3284\textwidth]{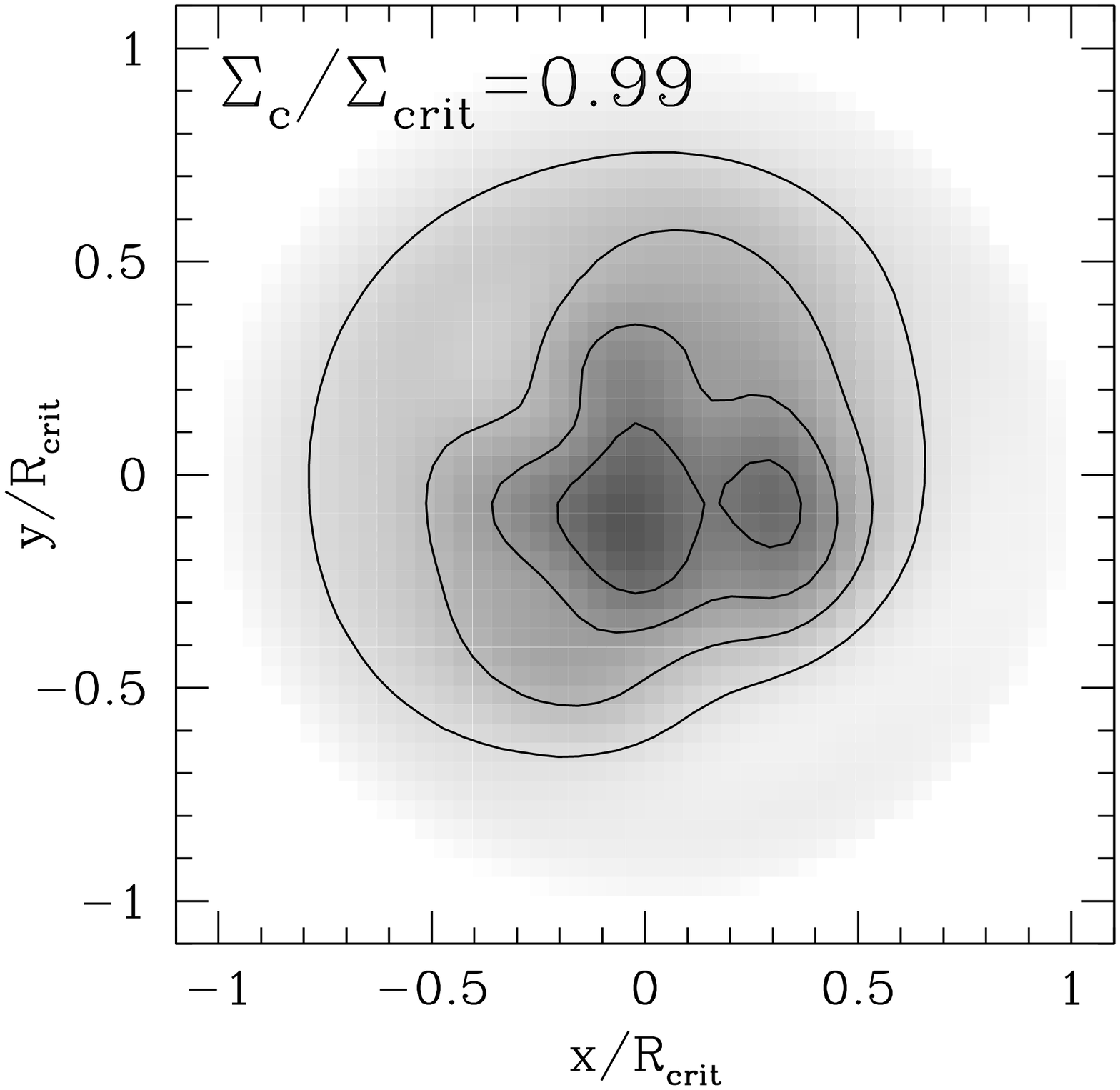}
\includegraphics[width=0.3284\textwidth]{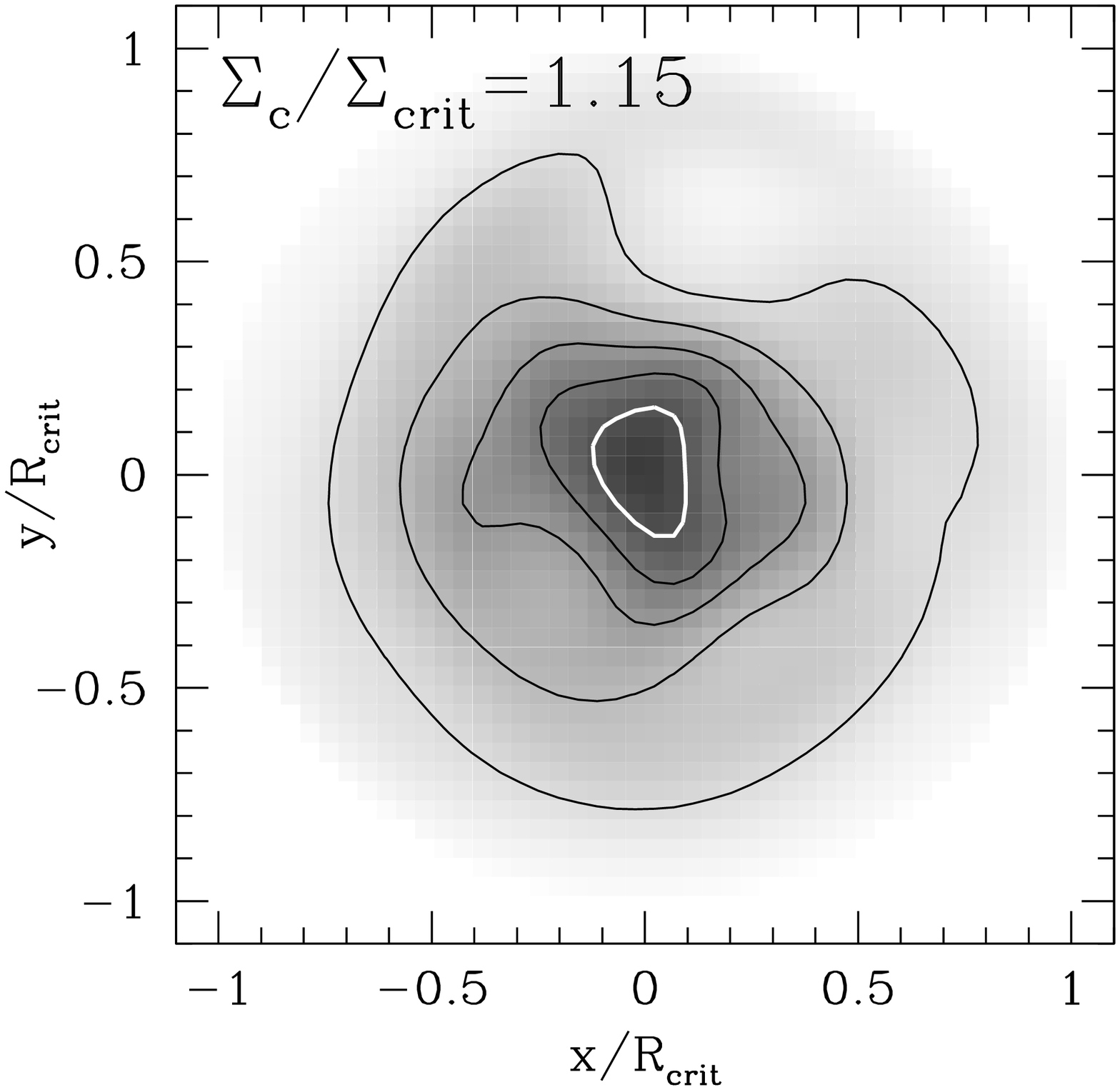}\\
\includegraphics[width=0.3284\textwidth]{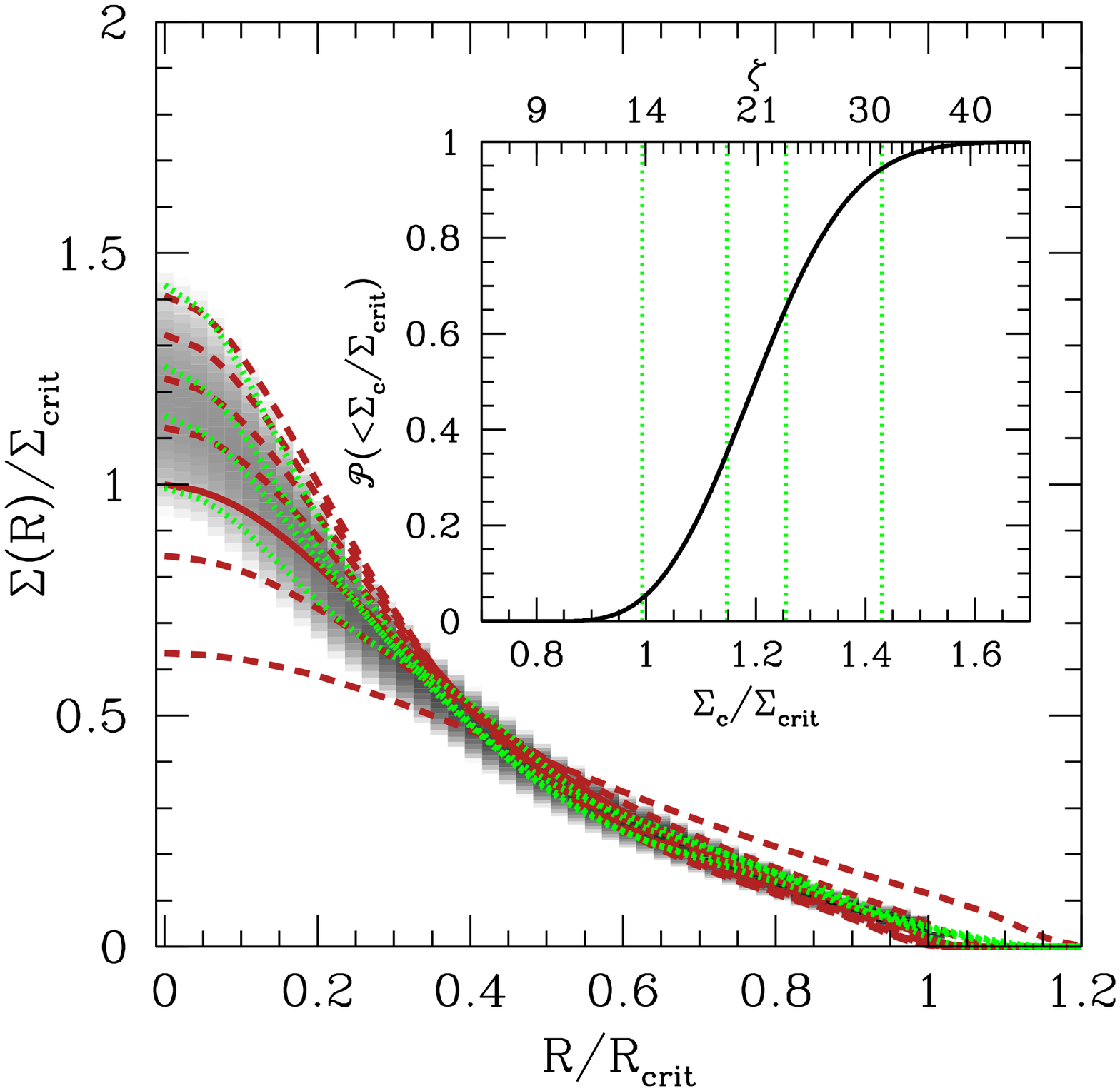}
\includegraphics[width=0.3284\textwidth]{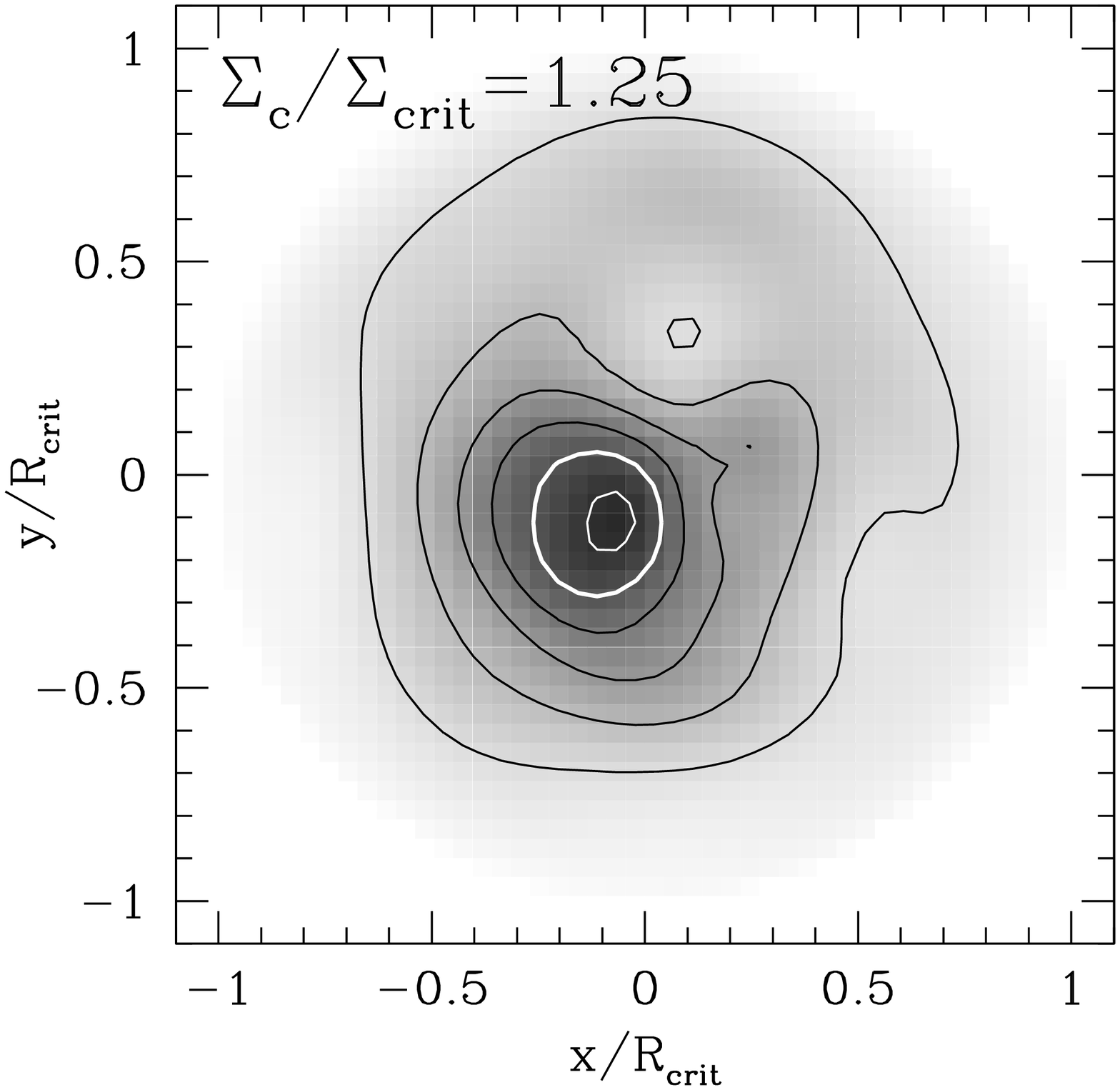}
\includegraphics[width=0.3284\textwidth]{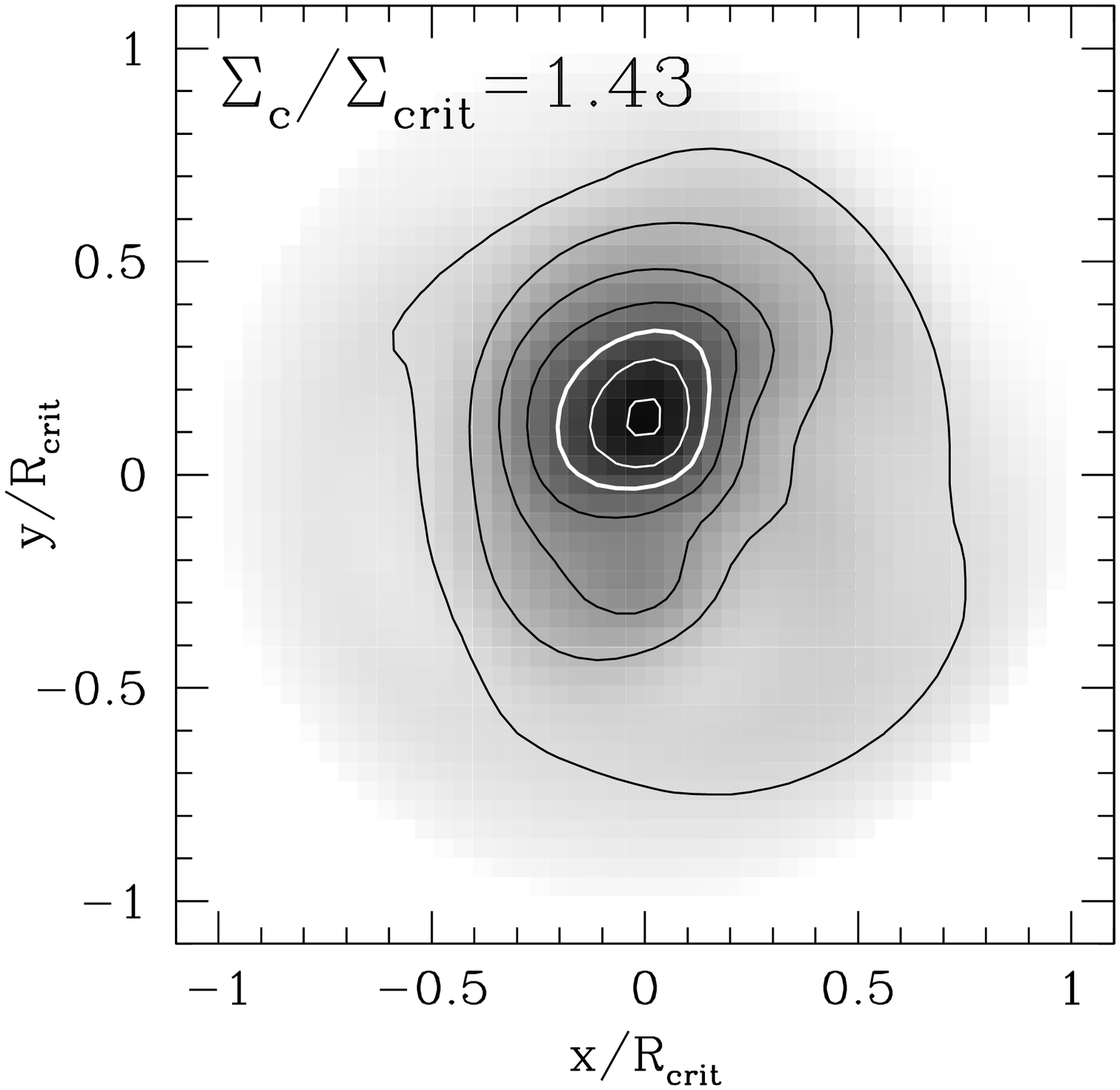}
\end{center}
\caption{Column density profiles and representative maps for pulsating
  cores with $\zeta=14$, $\Eo=0.3\Eb$ and a Kolmogorov energy
  spectrum (Equation \ref{eq:KES}).  Upper Left: Probability distribution of the
  azimuthally-averaged column density distribution, normalized by the
  central column density of the critical isothermal core of the same
  mass.  The gray-scale is logarithmic, with the floor at a
  probability density of 0.01.  The inset shows the probability
  distribution of the central column density.  For reference the
  central column density of the static configuration is also shown.
  On the top axis the $\zeta$ corresponding to a static isothermal
  core with the same mass is shown.  Bottom Left: Isothermal core
  profiles (dashed red lines, from bottom to top, $\zeta=6$,
  $10$, $18$, $22$, $26$ \& $30$; and solid red line, $\zeta=14.04$)
  and illustrative core profiles (green dotted lines) for cores with
  the central column densities at which the cumulative probability
  equals 5\%, 35\%, 65\% \& 95\%, shown explicitly in the inset.
  Center \& Right: Example column density maps, corresponding to the
  profiles shown in the lower left panel.  The gray-scale ranges from
  $\Sigma/\Sc=0$ to $1.5$, and subsequent contours are separated in
  steps of 0.2, with black contours for values less than unity and
  white above.
  \label{fig:KetoCasellids}}
\end{figure*}

\begin{figure*}
\begin{center}
\includegraphics[width=0.3284\textwidth]{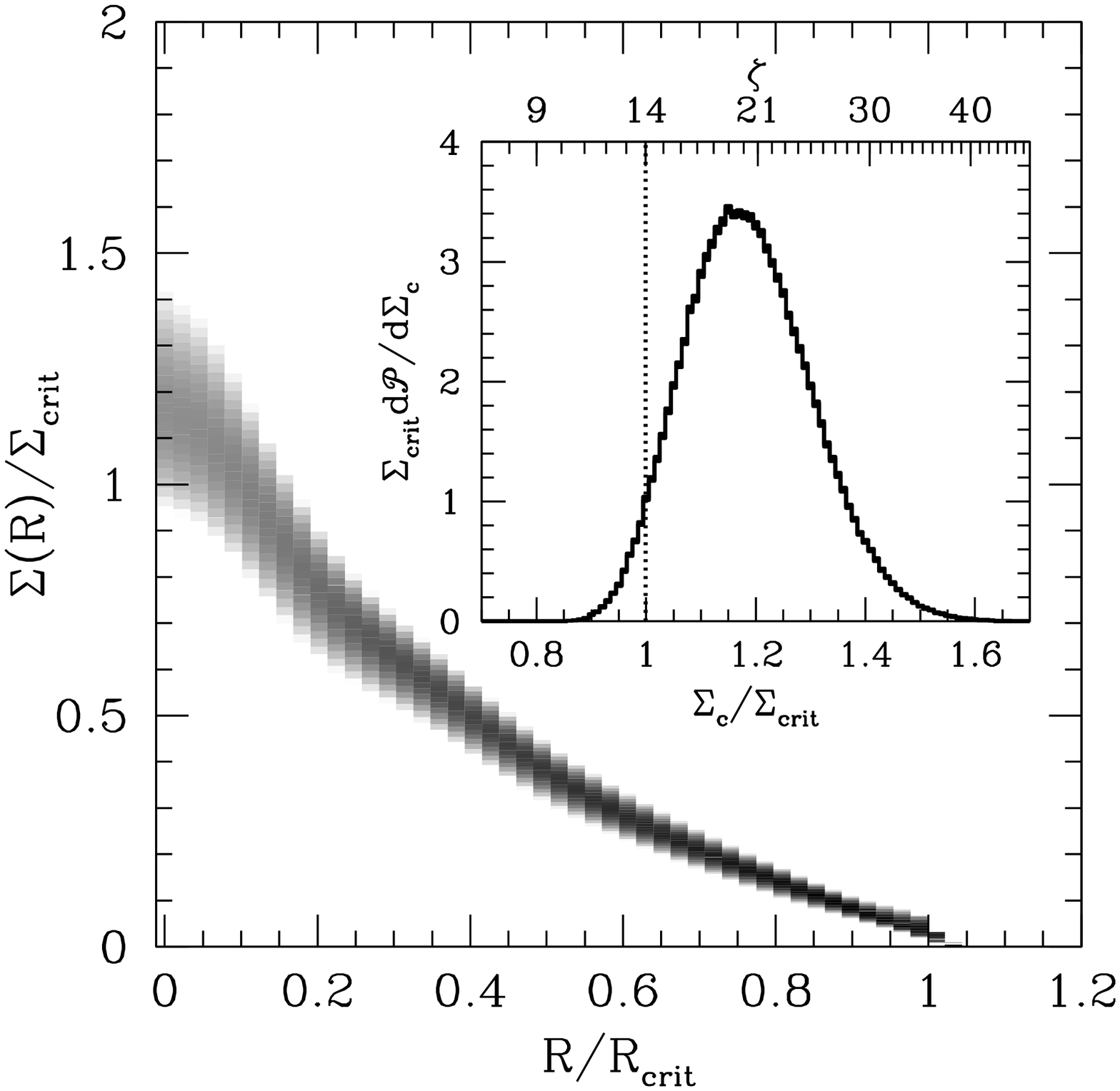}
\includegraphics[width=0.3284\textwidth]{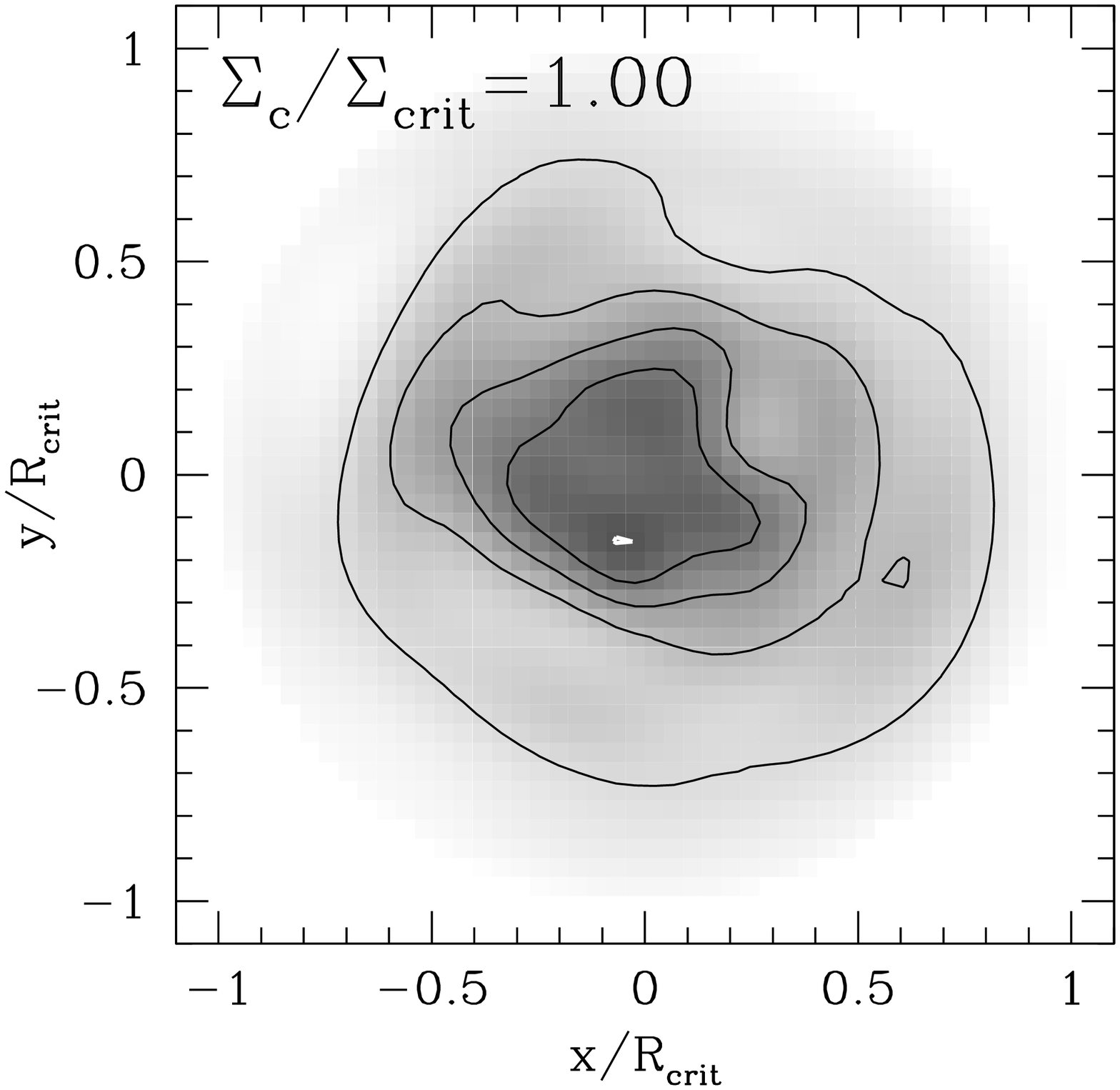}
\includegraphics[width=0.3284\textwidth]{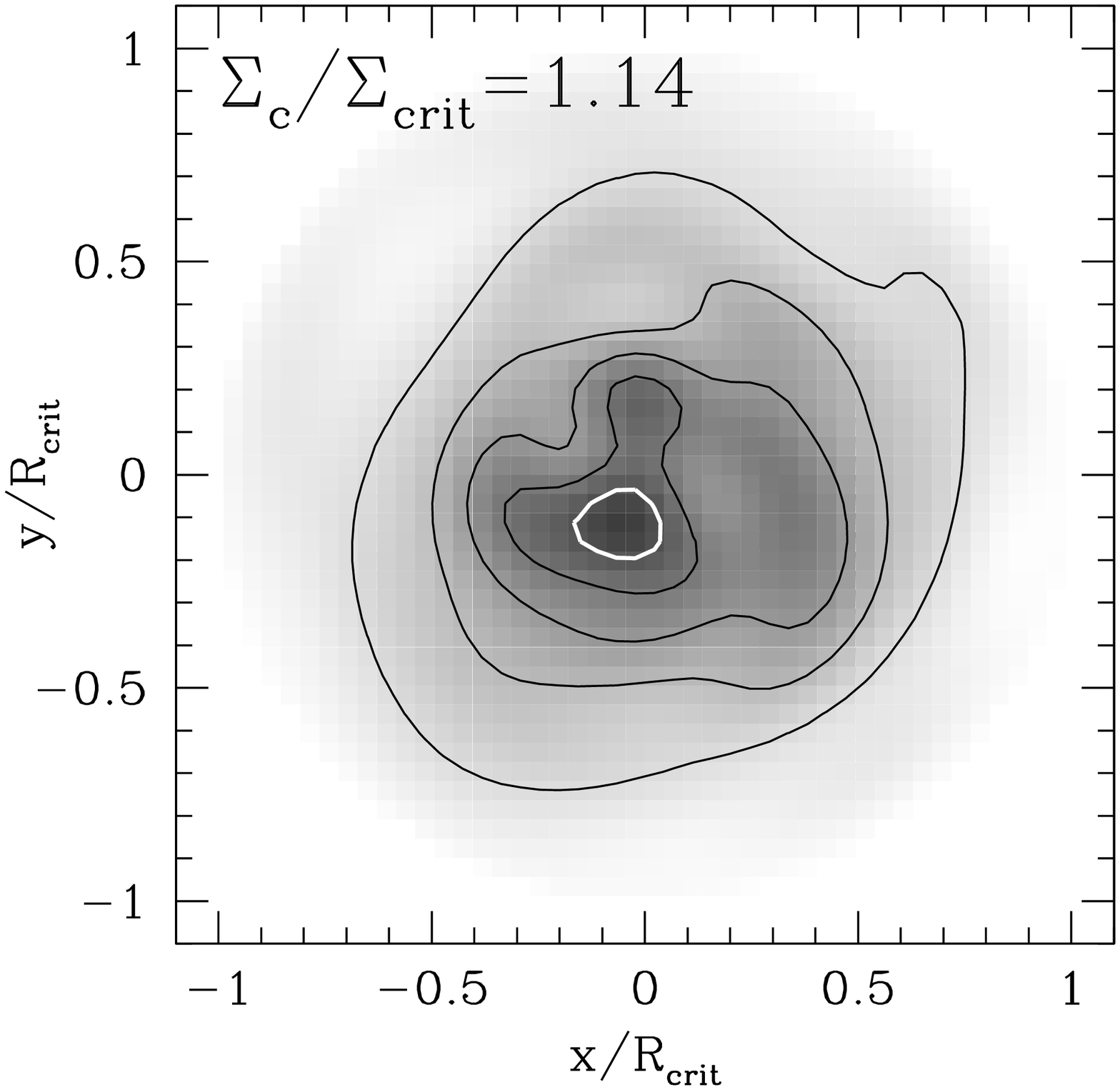}\\
\includegraphics[width=0.3284\textwidth]{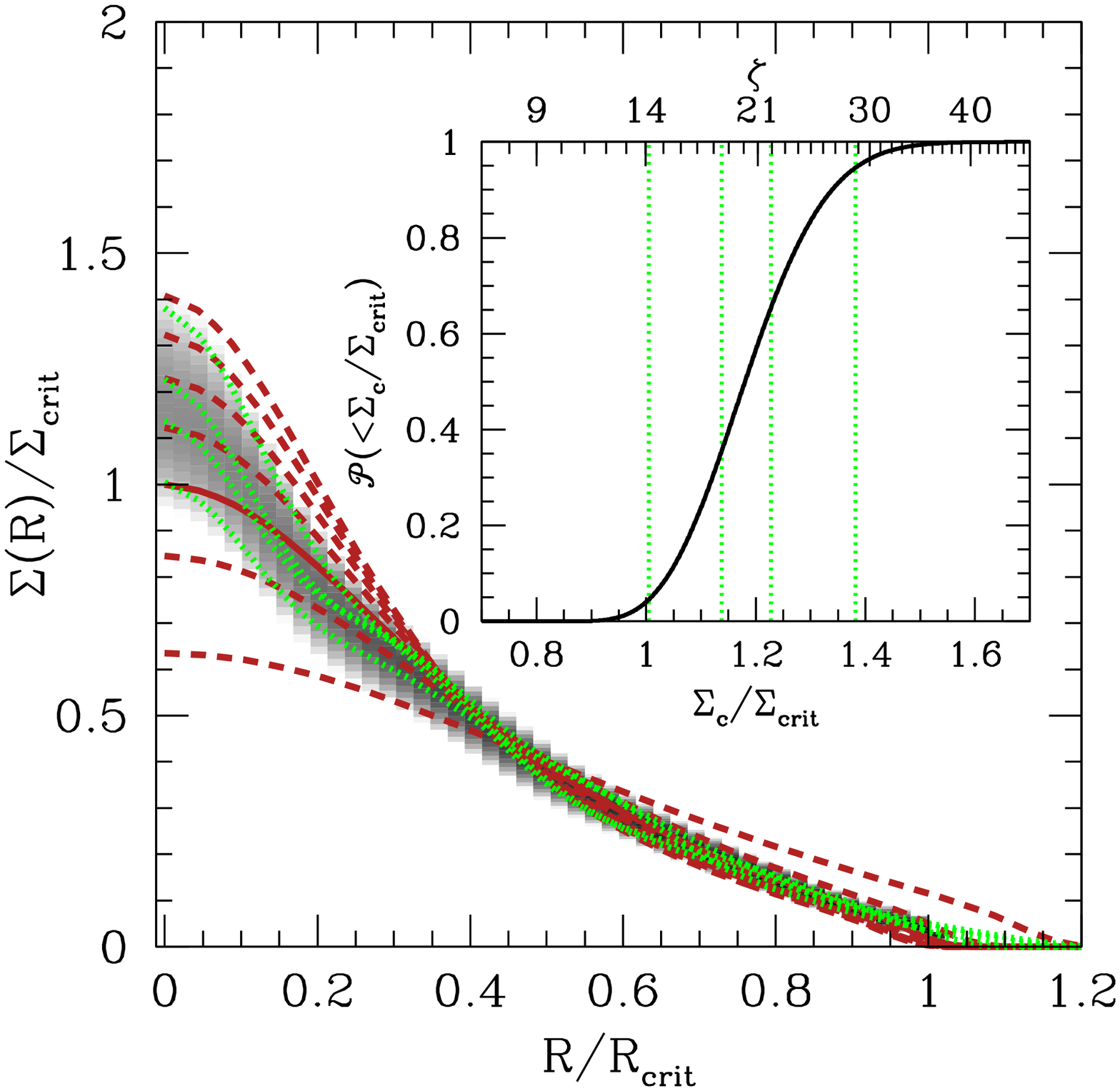}
\includegraphics[width=0.3284\textwidth]{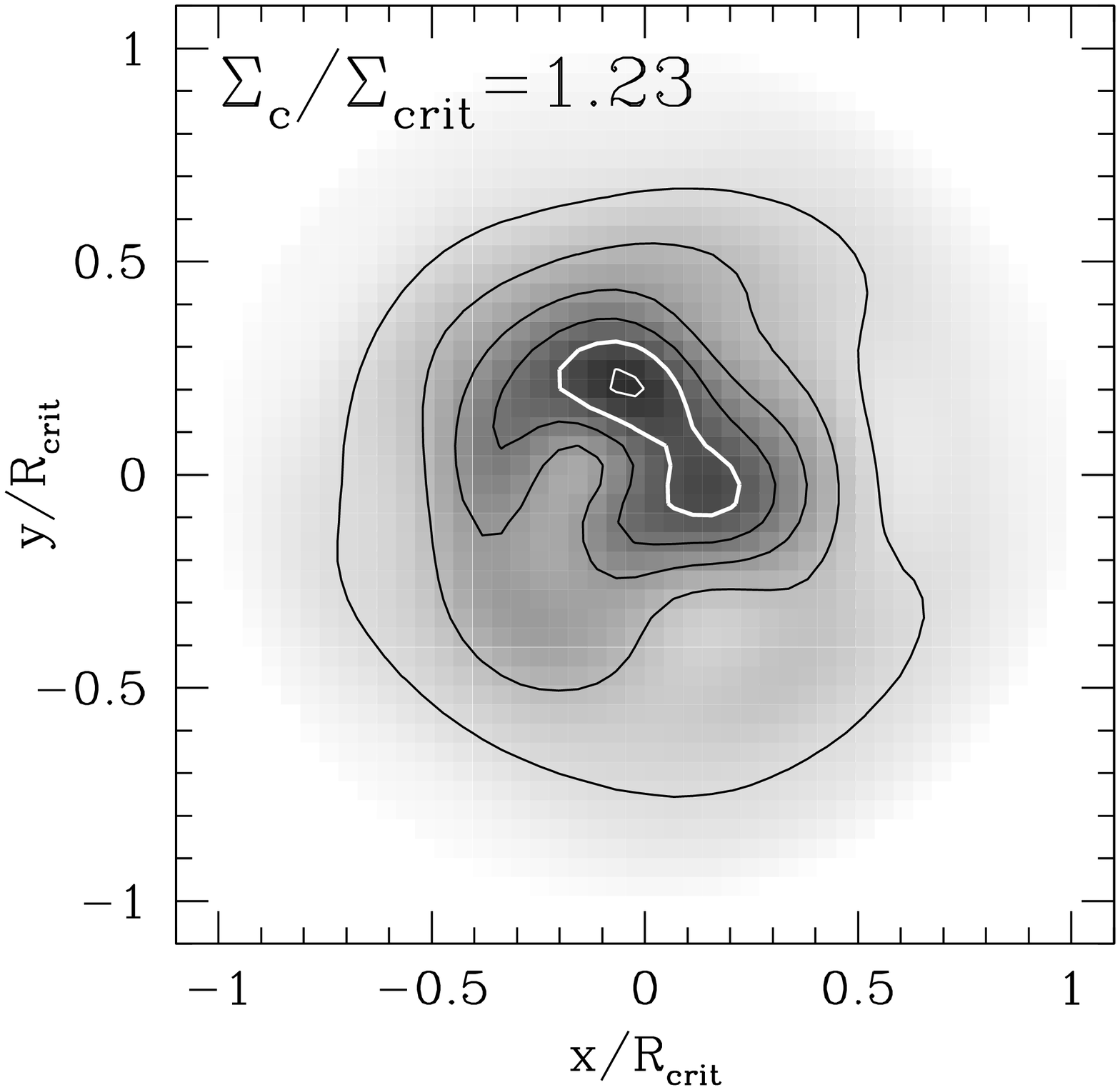}
\includegraphics[width=0.3284\textwidth]{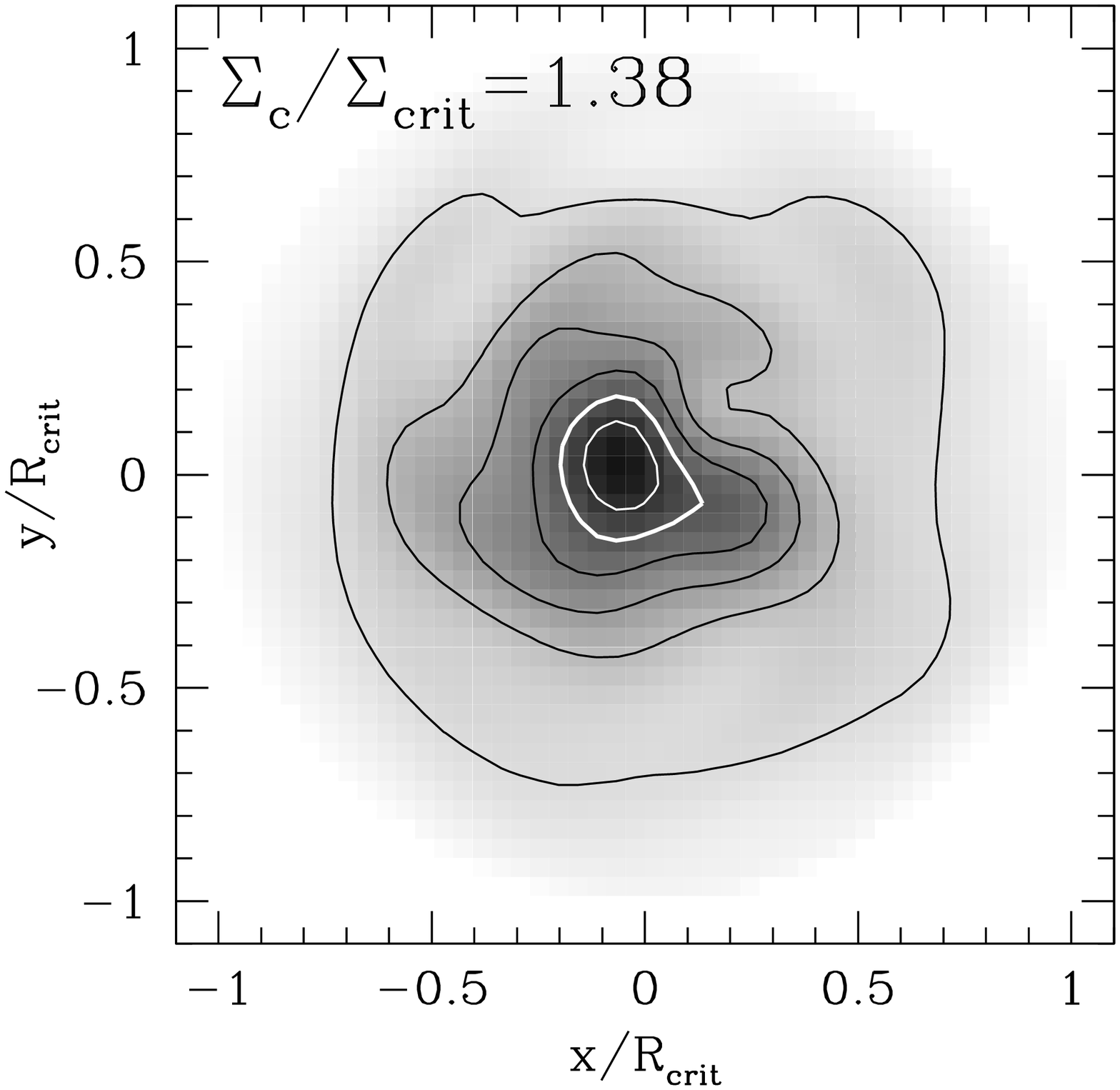}
\end{center}
\caption{Column density profiles and representative maps for pulsating
  cores with $\zeta=14$, $\Eo=0.3\Eb$ and a Flat energy
  spectrum (Equation \ref{eq:FES}).  For further details see Figure \ref{fig:KetoCasellids}.
  \label{fig:Fcds}}
\end{figure*}

\begin{figure*}
\begin{center}
\includegraphics[width=0.3284\textwidth]{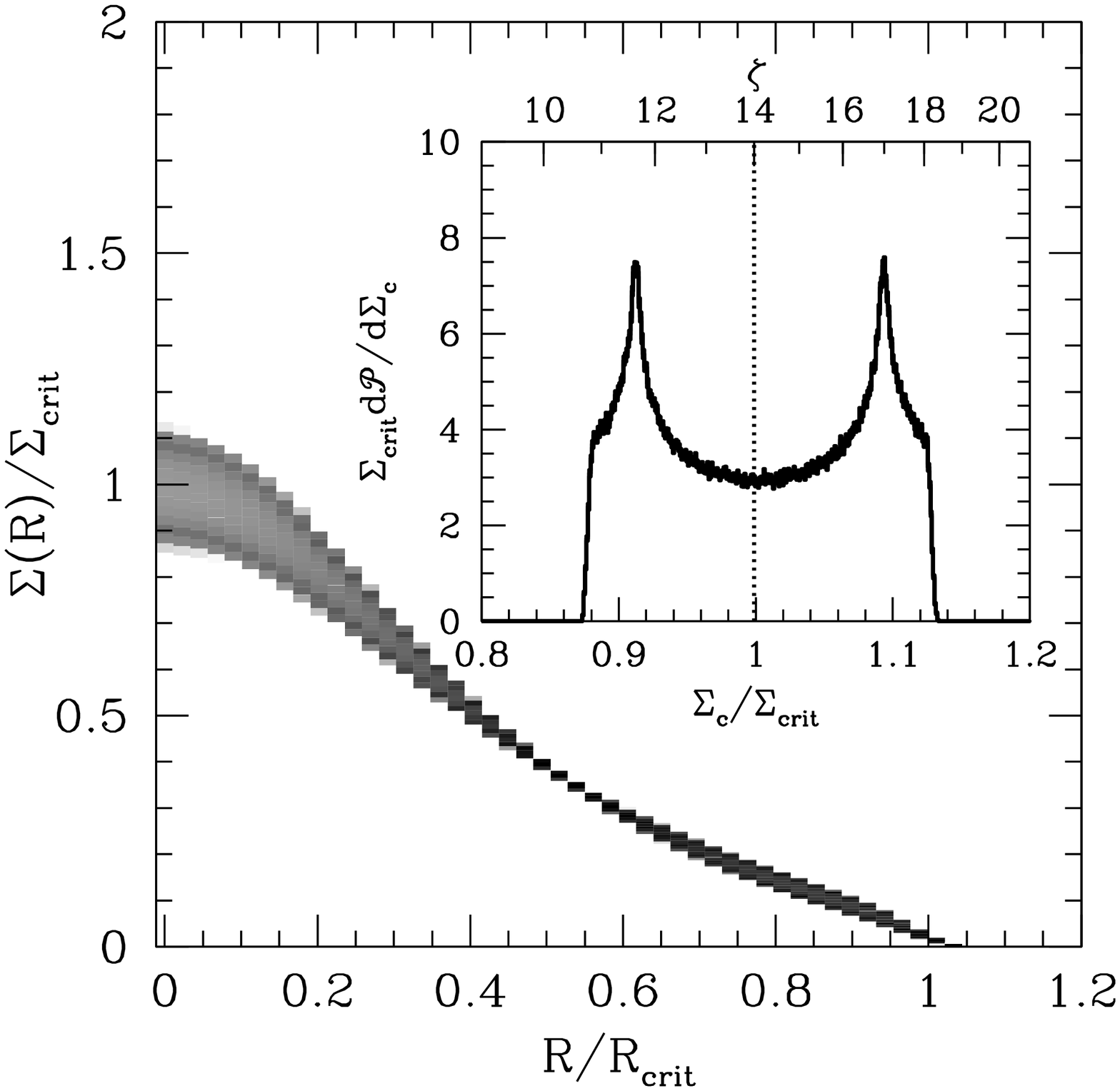}
\includegraphics[width=0.3284\textwidth]{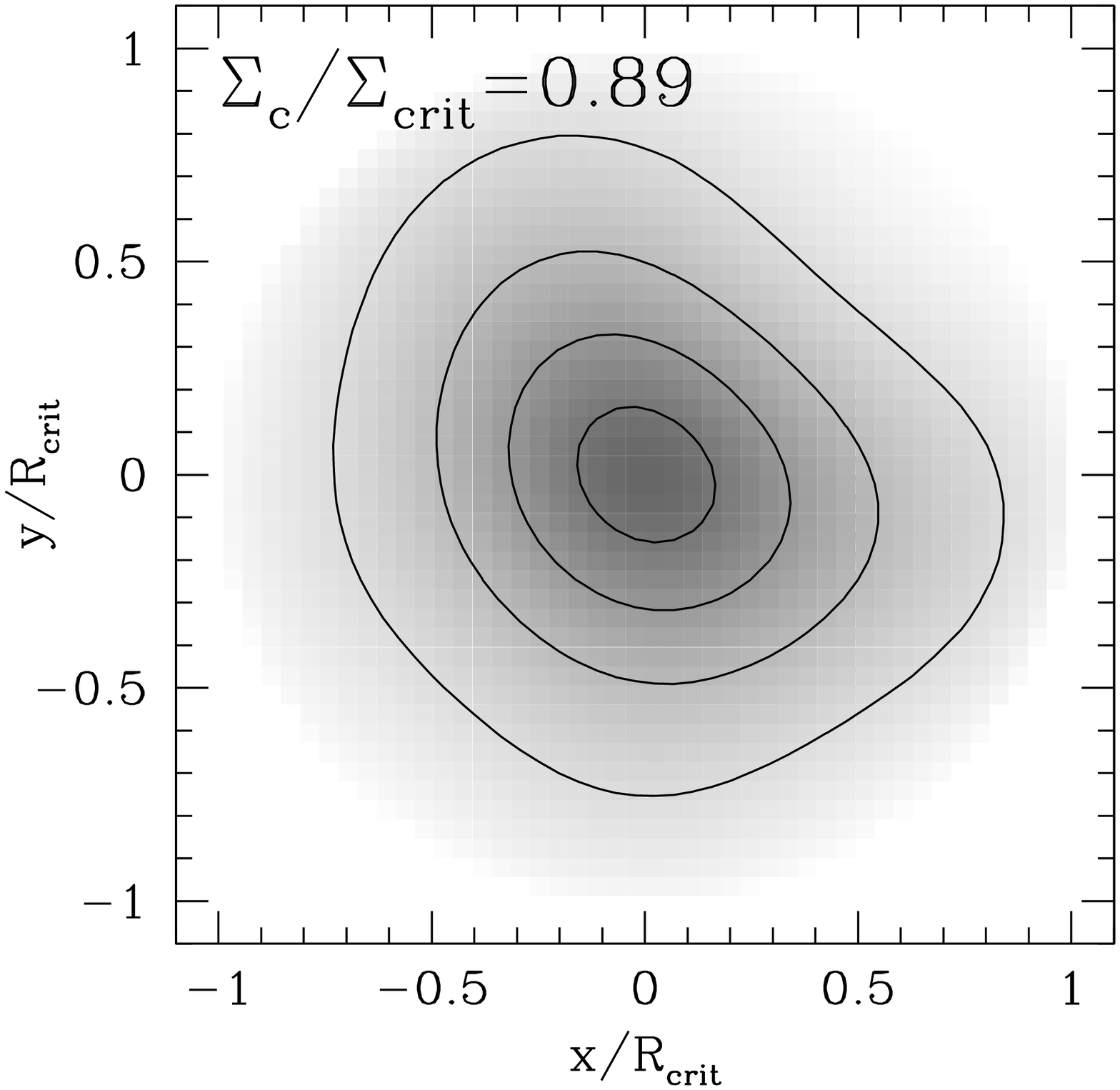}
\includegraphics[width=0.3284\textwidth]{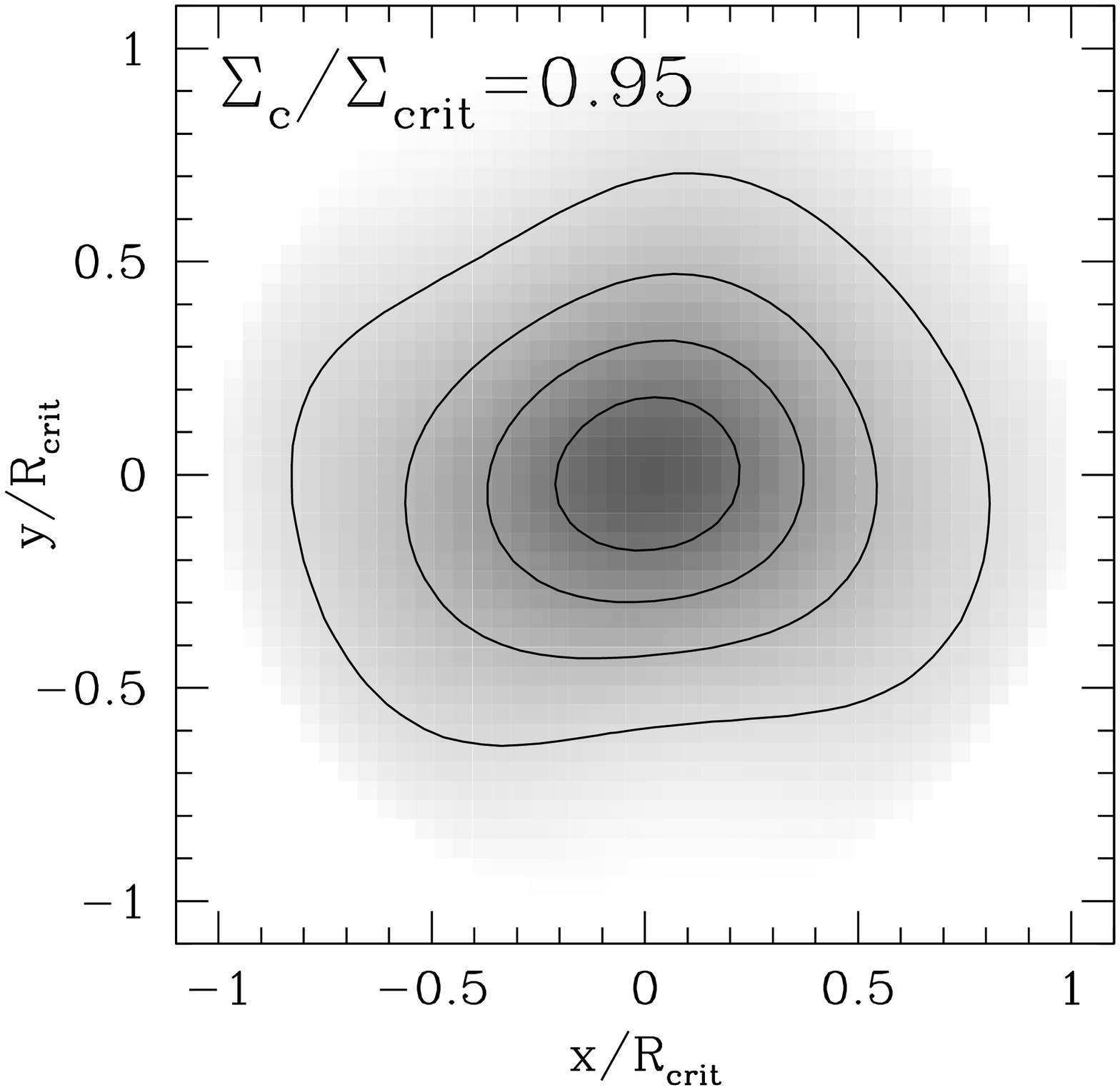}\\
\includegraphics[width=0.3284\textwidth]{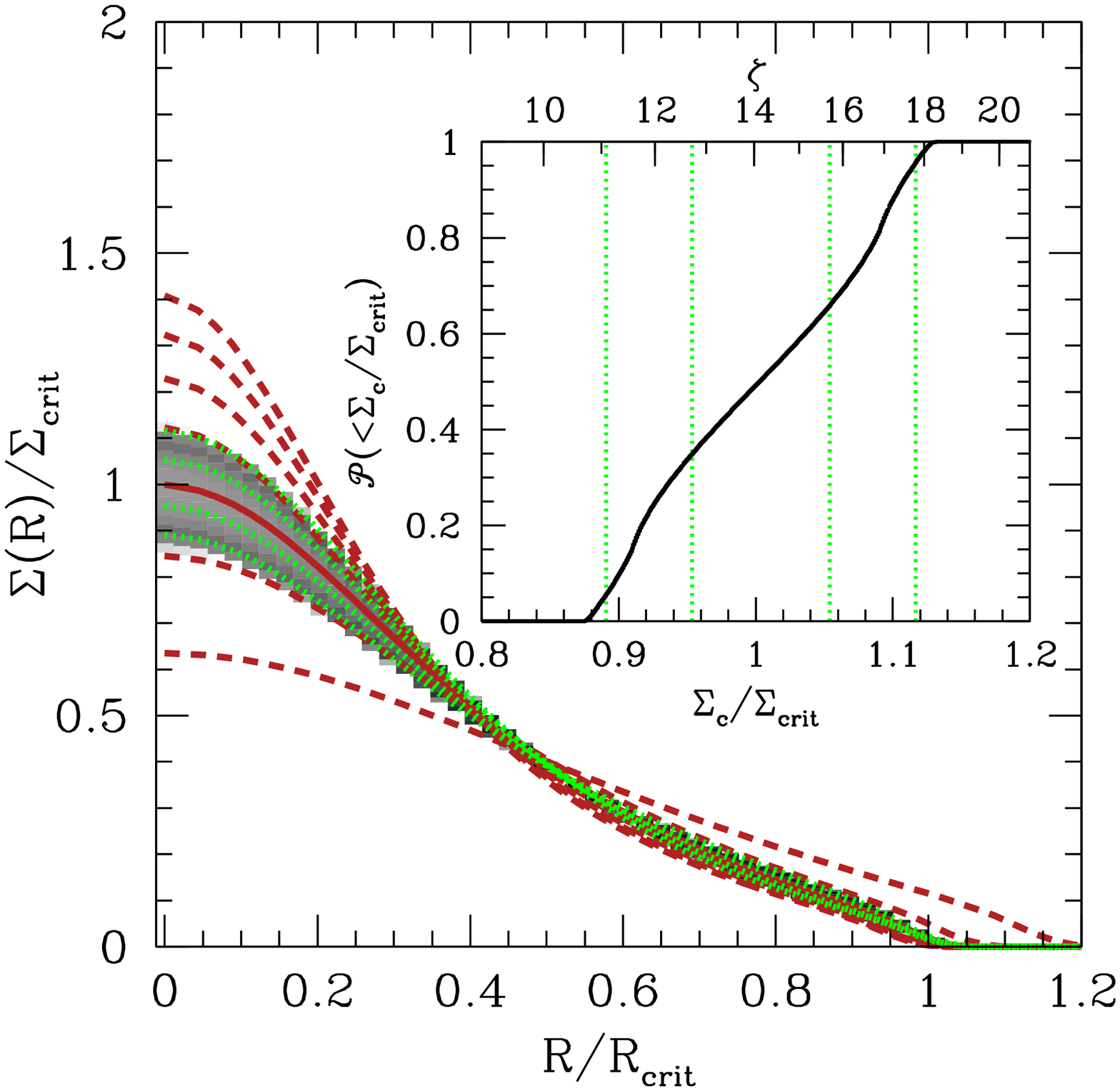}
\includegraphics[width=0.3284\textwidth]{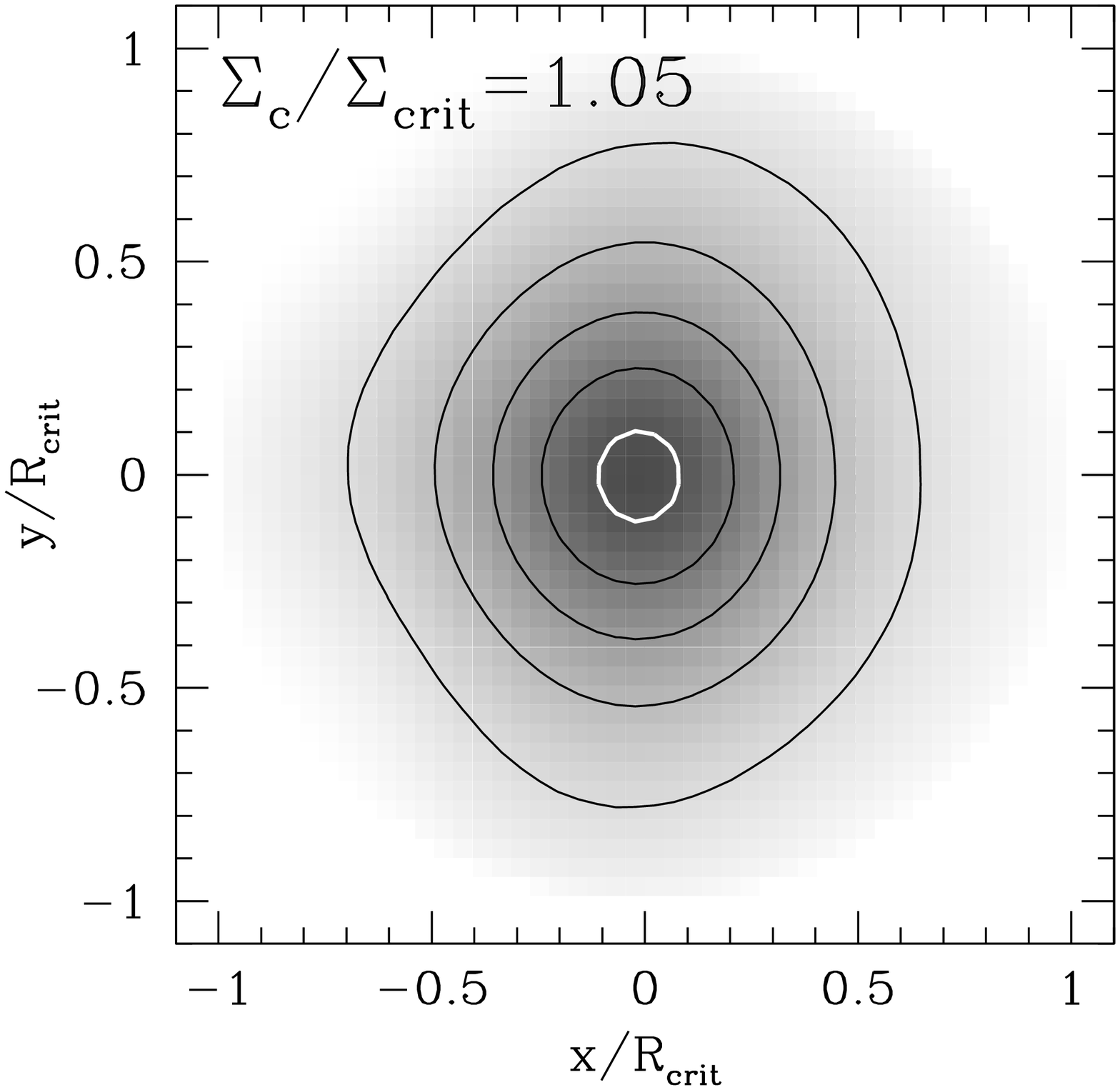}
\includegraphics[width=0.3284\textwidth]{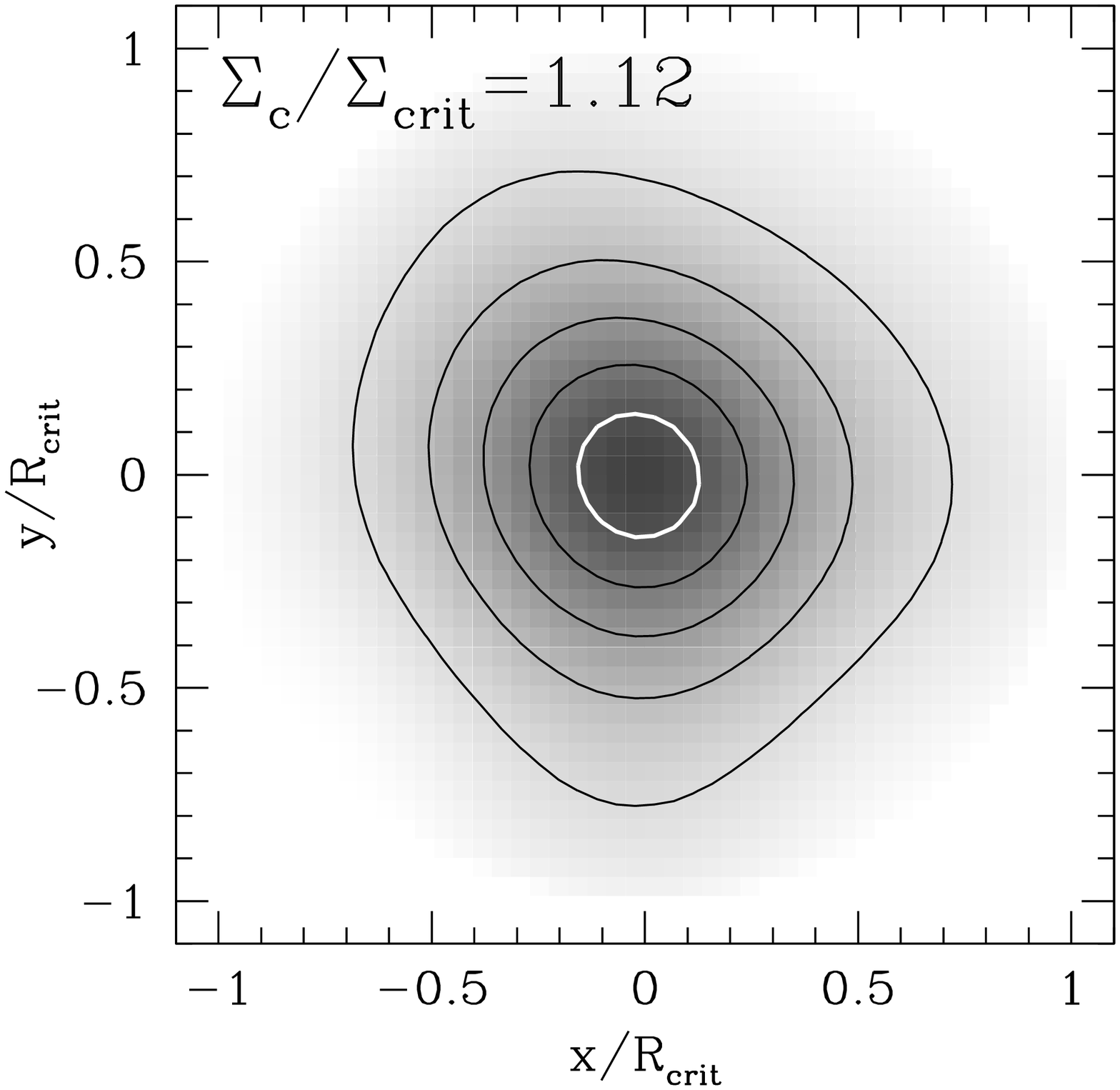}
\end{center}
\caption{Column density profiles and representative maps for pulsating
  cores with $\zeta=14$, $\Eo=0.3\Eb$ and a fundamental-Kolmogorov
  energy spectrum (Equation \ref{eq:fKES}).  Note the change in scale in the left-panel
  insets.  For further details see Figure \ref{fig:KetoCasellids}.
  \label{fig:fcds}}
\end{figure*}

While not motivated by any particular turbulence model, a
flat energy spectrum does have the virtue that all realistic choices
for the turbulent spectra lie between it and the Kolmogorov
distributions. Therefore, we also consider,
\begin{equation}
E_{nlm}^F\propto
\left\{\begin{array}{ll}
\displaystyle k_{nlm}^{-2} & \quad {\rm if~}n\le5~{\rm and}~0<l\le5\\
0 & \quad{\rm otherwise,}
\end{array}
\right.
\label{eq:FES}
\end{equation}
which is flat for $n,l\le5$ and vanishing otherwise.  The truncation
is arbitrary, here set by numerical convenience.  However, as argued
above, the particulars of the cut-off are not important in practice
since it is primarily the relative values of $\Eo^0$ and $\Eo^{>0}$
that matter.  As with the Kolmogorov spectrum the latter dominates
with $\Eo^{>0}/\Eo^0\simeq 3.9$ implying that the resulting
perturbation to the surface pressure will be similar.  Indeed,
despite the considerable difference in their particulars, since
$\Eo^{>0}/\Eo^0$ are similar for both spectra the resulting surface
pressures are nearly indistinguishable. The surface pressure for the
Flat energy spectrum is shown in Figure \ref{fig:PvR} by the blue
line (when $\Eo=0.3\Eb$).  For both cases we find $\Po/P_0\simeq 0.5
\left|\Eo/\Eb\right|$, implying that the fractional change in the
maximum stable mass of an oscillating isothermal core is approximately
$0.25\left|\Eo/\Eb\right|$. 

Finally, we consider a fundamental-dominated energy spectrum, i.e., one
in which only the $n=0$ modes are excited, with an otherwise Kolmogorov
spectrum, as described for $E_{nlm}^K$.  That is,
\begin{equation}
E_{nlm}^{fK}\propto
\left\{\begin{array}{ll}
\displaystyle k_{nlm}^{-11/3} & \quad {\rm if~}n=0~{\rm and}~l>0\\
0 & \quad{\rm otherwise}
\end{array}
\right.
\label{eq:fKES}
\end{equation}
Aside from presenting an extreme alternative, such an energy spectrum
will be motivated in Section \ref{sec:EoaMSSC}, where we discuss the
evolution of the energy spectrum during core formation.  In this case
$\Eo=\Eo^0$ explicitly, and the resulting surface pressure, shown by
the green line in Figure \ref{fig:PvR} (again for $\Eo=0.3\Eb$), is
substantially larger than those produced by the previous two spectra.
This is a result of both the dependence of $\Po$ upon the division of
energy between $\Eo^0$ and $\Eo^{>0}$ as well as the departure from
the asymptotic expressions of the
$\omega_{nlm}$ near $\zc$ for $n>0$.  As a consequence, in this case
we find $\Po/P_0\simeq 1.0 \left|\Eo/\Eb\right|$, corresponding to a
fractional increase in the maximum stable mass of approximately
$0.5\left|\Eo/\Eb\right|$.

In summary, we have found that the non-radial oscillations generally
produce an increase in the surface pressure of the isothermal cores,
with magnitude linearly related to the total energy in the pulsations
and bounded near $\zc$ by
\begin{equation}
0.5\left|\frac{\Eo}{\Eb}\right|
\lesssim \frac{\Po}{P_{\rm crit}} \lesssim
1.0\left|\frac{\Eo}{\Eb}\right|\,,
\end{equation}
for an extraordinarily broad class of energy spectra.  This
corresponds to an increase in the maximum mass of stable cores of
approximately
\begin{equation}
0.25\left|\frac{\Eo}{\Eb}\right|
\lesssim \frac{\Delta M}{M_{\rm crit}} \lesssim
0.5\left|\frac{\Eo}{\Eb}\right|\,.
\end{equation}

\section{The Structure of Mode-Supported\\ Equilibrium Configurations}\label{sec:MSEC}
Thus far we have characterized the effects of oscillations in terms of
the unperturbed core parameters.  However, large-amplitude pulsations
can dramatically alter the observed appearance of a core, with important
consequences for the interpretation of starless core statistics.
In \citet{KetoField2005}, \citet{Keto2006}, and \citet{Broderick2007}
we showed how pulsations alter the molecular line profiles and widths.
Here we explore how pulsations modify 
the apparent core column density profiles, $\Sigma(R)$.  In this
section we study the range of perturbed $\Sigma(R)$, finding that
generally these are substantially asymmetric, frequently devolving
into many peaked structures, and are biased towards
azimuthally-averaged profiles that are best fit with super-critical
static isothermal configurations.  That is, stable, pulsating starless
cores will typically appear to be strongly asymmetric and super-critical
if fit with a static BE model.

\subsection{Generating Column Densities of Pulsating Cores}
To ascertain the typical observational consequences of the
oscillating cores, we generate an ensemble of pulsating cores.  We do
this for each of the energy spectra described in Section
\ref{sec:TOES} by setting the density to
\begin{equation}
\rho(\bmath{r})
=
\rho_0(\bmath{r})
+
\Re\left[
\sum_{nlm}
\sqrt{\frac{2 E_{nlm}}{M_{nlm} \omega_{nlm}^2}}
\e^{i\Phi_{nlm}} \rho_{nl}(r) Y_{lm}(\bmath{\hat{r}})
\right]\,,
\end{equation}
where the $\rho_0(\bmath{r})$ is the unperturbed density,
$\rho_{nl}(r) Y_{lm}(\bmath{\hat{r}})$ are the density perturbation
eigenfunctions and the $\Phi_{nlm}$ are randomly determined phases,
corresponding to a random realization of the particular oscillations.
Any realization for which $\rho(\bmath{r})$ vanishes anywhere is
discarded, though this has little effect upon the statistical
quantities since the distribution of the various observables is nearly
identical for the accepted and discarded cores.  

We do not vary the $E_{nlm}$, nor do we consider different underlying
equilibrium configurations, choosing for concreteness a stable configuration,
near critical,
with $\zeta=14$.  Since we vary only the phases, $\Phi_{nlm}$,
we are not producing a complete ensemble; 
such an ensemble would necessarily have variations
in the amplitudes of the oscillations, the total oscillation energy
and the core masses, in addition to the mode phases.  Rather, in the
absence of strong constraints upon those quantities,  we restrict
ourselves to highlighting the consequences of the pulsations upon the
measured densities with fixed amplitudes and core masses.

Once we have generated a $\rho(\bmath{r})$ we construct the column
density map, $\Sigma(\bmath{R})$, by integrating along the z-axis.
These are normalized to the central column density of the static,
critical configuration, $\Sc$.  While the center of the
unperturbed core is always located at the origin, we define the
``central'' column density, $\Sigma_c$, to be
$\max\left[\Sigma(\bmath{R})\right]$, corresponding to how the center
of the core would be defined in practice.  Thus, configurations which
have $\Sigma_c/\Sc>1$ appear to be super-critical.

Finally, we produce azimuthally-averaged column density profiles,
$\Sigma(R)$, by averaging in concentric annuli about the peak column
density.  Since the perturbations can both raise and lower
$\Sigma(\bmath{R})$, this observationally motivated definition of
$\Sigma_c$ necessarily biases us towards higher column densities, and
thus towards identifying pulsating cores as closer to super-critical
than the are.

We repeat this process many times, building up a statistical
description of the pulsating core column densities.  The results are
summarized for $\Eo=0.3\Eb$ in Figures \ref{fig:KetoCasellids}--\ref{fig:fcds},
and are discussed in detail below.

\subsection{Central Column Densities}
The probability of measuring a given central column density is shown
in the inset of the upper-left panel of Figures
\ref{fig:KetoCasellids}--\ref{fig:fcds}.  In all cases this
probability is broadly distributed about $\Sigma_c/\Sc=1$.  As with
their dynamical consequences, the distributions arising from
pulsations with Kolmogorov ($E_{nlm}^K$) 
and Flat ($E_{nlm}^F$) energy spectra are quite similar. Both probability
distributions are
centrally peaked at a value significantly larger than unity,
$\Sigma_c/\Sc\simeq1.2$, corresponding to a $\zeta\simeq21$, with only
a roughly 5\% probability of finding $\zeta\le\zc$ (seen in the
cumulative probability distribution in the inset of the lower-right
panels of Figures \ref{fig:KetoCasellids}--\ref{fig:fcds}).  Furthermore, these
can appear to be extraordinarily super-critical, exhibiting $\zeta$'s
of up to 28 for $\Eo=0.3\Eb$.

Differing from these qualitatively, the distribution arising from the
fundamental-Kolmogorov energy spectrum ($E_{nlm}^{fK}$) is double
peaked.  This is due to the dominance of the fundamental quadrupole
mode (the dipole has no fundamental).  It is also nearly symmetrically
distributed about $\Sigma_c/\Sc=1$, somewhat more narrowly than with the
Kolmogorov and Flat spectra, and with ${\mathscr P}(\zeta\le\zc)\simeq50\%$.
Despite this, the double peaked nature of the distribution implies
that most cores will appear to have $\zeta$'s significantly larger
or smaller than $\zc$, reaching up to roughly $17$ for $\Eo=0.3\Eb$.

\subsection{Column Density Profiles}
The results from the central column densities are borne out by the
azimuthally averaged column density profiles.  As before, the
Kolmogorov and Flat energy spectra produce quantitatively similar
results.  The probability distributions of $\Sigma(R)/\Sc$ are shown
in the leftmost two panels in Figures \ref{fig:KetoCasellids}--\ref{fig:fcds}.
Over-plotted in red within the bottom left panels are a number of static
isothermal core profiles, for a number of $\zeta$.  $\Sigma(R)$ for
the critical case is shown by the solid line, with profiles moving
upward and downward in $\zeta$ increments of $4$.  In particular, the
$\zeta=18$ and $22$ profiles do a fair job of fitting the averaged
column density profiles, while those with significantly smaller
$\zeta$ do not.  For reference, profiles associated with randomly
chosen examples with central densities at the 5\%, 35\%, 65\% and 95\%
cumulative probabilities are shown by the green dotted lines (in each
Figure, these are associated with the maps shown on the right).

The fundamental-Kolmogorov energy spectrum produces similar column
density profiles.  As anticipated by the $\Sigma_c/\Sc$ distribution,
these are centered upon the critical profile, with both low and
high-$\zeta$ profiles represented.  In this case the averaged profiles
appear to fit the static-core profiles more accurately, though this
may be due to the decreased deviation from the critical core profile.

This suggests that if observations of pulsating cores are compared
to a static BE model, the observed cores will quite commonly appear
strongly super-critical with $\zeta > \zeta_{crit}$ even though the cores
are in no imminent danger of collapse.

\subsection{Column Density Maps}

Despite the approximate equivalence between the average column density
profiles of stable pulsating cores and static super-critical cores, the full
two-dimensional maps of pulsating cores can be dramatically asymmetric.  Example column
density maps, randomly chosen, are shown in the center and right
columns of Figures \ref{fig:KetoCasellids}--\ref{fig:fcds}, corresponding to
the same configurations for which the azimuthally averaged
distributions are shown in the bottom left panel of each Figure.
These are representative of the configurations appearing at the 5\%,
35\%, 65\% and 95\% cumulative probabilities, giving some idea of how
typical and, perhaps, rare cores appear.

Again the column density maps associated with the Kolmogorov and Flat
energy spectra are similar, though maps associated with the former are
typically more centrally concentrated.  In roughly 20\% and 10\% of
cases, for the Kolmogorov and Flat spectra, respectively, the column
density maps are double-peaked.

With fewer degrees of freedom, and smaller net density variations for
the same $\Eo$, it is not surprising that the fundamental-Kolmogorov
spectrum shows less small-scale structure.  In most cases the column
density maps are only weakly non-circular.  Driving $\Eo$ up to
$0.6\Eb$ produces roughly comparable variations in $\Sigma_c/\Sc$, and
correspondingly larger deviations from cylindrical symmetry, typically
with aspect ratios larger than 2.

Nevertheless, for all three energy spectra considered it is possible to
produce manifestly asymmetric structures, comparable to those observed.  
As a consequence, strong departures from symmetry in the
column densities are not generally evidence for unstable structures.

\section{Evolution of Mode-Supported Starless Cores}\label{sec:EoaMSSC}
In addition to the excitation mechanism, the evolution of the
underlying core properties and the non-linear damping of oscillations
can imprint themselves upon the pulsation energy spectrum.
Ultimately, the relative importance depends upon the relationship
between the timescale associated with each process.  Here we present
some observationally motivated constraints  and discuss the
consequences for the stability and appearance of starless cores.  In
particular we motivate a picture in which the oscillations are inherited
from the parent molecular cloud turbulence, and the fundamental
modes are amplified at the expense of shorter wave length  during a slow
contraction phase resulting in a fundamental-mode dominated spectrum.
Subsequent non-linear decay of the pulsations then eliminates the
additional turbulent support leading to collapse and star formation
in those cores that are super-critical with respect to the static BE model.

\subsection{Timescales}

The relevant timescales for the oscillating starless cores
are the evolution timescale of the core, $T$, the
pulsation periods, $P_{nlm}$, lifetimes, $\tau_{nlm}$, and excitation
timescale $t_e$.  Even without knowing the excitation mechanism we
can begin to place these into a hierarchy based upon the observation
that cores with large-amplitude, long-period pulsations exist.  
In addition to the observations of \citet{Lada2003},
\citet{Redman2006} and \citet{Aguti2007} mentioned in the
introduction, the observations of \citet{Sohn2007} find that about 2/3
of all their 85 observed cores show clear evidence of either expansion
or contraction. Most of the remaining 1/3 show complex spectral line
profiles that also indicate internal motions but are difficult to
characterize.  The presence of coherent oscillations immediately
implies that
$P_{nlm}<T,\,\tau_{nlm},\,t_e$.  That is, for coherent oscillations to
exist the core must persist in a fixed state sufficiently long for the
pulsation to coordinate the large-scale motions throughout the core,
and the pulsation amplitudes cannot change significantly during this
process.  Violations of either of these would result in motions that
appear stochastic (microturbulent) instead of coherent.

That the observed pulsations have $\Eo\sim\Eb$ implies
$\tau_{nlm}\lesssim t_e$.  Were the observed modes the result of a prolonged
excitation, or many individual excitation events, we would expect with
near unit probability that a given core will have had a sequence of
excitations resulting in $\Eo\gg\Eb$, at which point the core is torn
apart, in direct conflict with the observation of numerous cores with
substantial internal motions.  In principle, this can be prevented by
limiting the energy transfer during each excitation event.  However,
in that case the development of large-scale pulsations is itself rare,
again in direct conflict with observations.  Thus, as long as the
excitation mechanism is uncorrelated with the pulsations themselves,
i.e., there is no oscillation ``feedback'', a given mode must have, on
average, originated from only a single excitation event.

Finally, we have $T\lesssim\tau_{nlm}$.  As the pulsations damp away,
the support arising from the oscillations is lost, forcing the core to
evolve towards higher central densities and then ultimately towards
collapse.  Thus, for cores in which the pulsations are dynamically
important the mode lifetimes set an upper limit upon the core
evolution time; a core may evolve on shorter timescales but it
{\em will} evolve as modes decay.

Thus, in summary, the very presence of starless cores with dynamically
significant pulsations gives the following hierarchy among the relevant
timescales:
\begin{equation}
P_{nlm} < T \lesssim \tau_{nlm} \lesssim t_e\,.
\end{equation}
This simplifies dramatically the determination of the consequences of
mode decay and core evolution.

\subsection{Energy Spectrum from Inherited Turbulence} \label{sec:ESfRT}

\begin{figure}
\begin{center}
\includegraphics[width=\columnwidth]{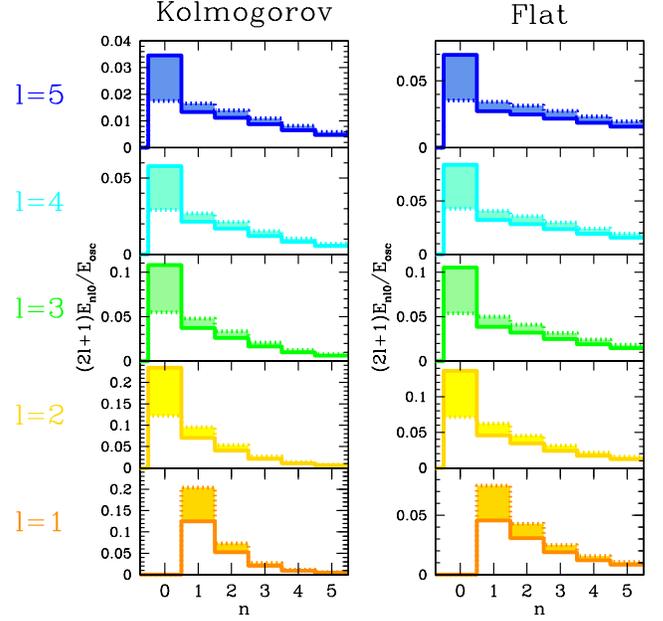}
\end{center}
\caption{Evolution of the pulsation energy spectra during adiabatic
  core collapse.  The dotted lines show $k_{nlm}^2 E_{nlm}/\Eo$ for
  the Kolmogorov (left) and Flat (right) energy spectra, assumed to be
  inherited at $\zeta=1.25$, as functions of $n$ for various $l$.  The
  solid lines show the adiabatically evolved spectra at $\zeta=14$,
  with in all cases the fundamental ($n=0$) modes gaining
  substantially more energy than the other oscillations.  In the case
  shown, by $\zeta=8$ the spectra have nearly attained their final
  state, implying that even for substantially sub-critical cores, the
  pulsation energy spectrum can be dominated by the fundamental modes.
  }\label{fig:Eevol}
\end{figure}

Given that $T\lesssim t_e$, we are justified in imagining that
pulsations on super-critical cores are relics of the core formation
process.  The natural mechanism for the excitation of the oscillations
is then the turbulence within the parent cloud, inherited at the
formation of the core.  This does not mean that the energy
spectrum of an observed core is always the same as the parent
cloud.   If cores are
formed by a prolonged collapse, the evolution of the underlying
equilibrium configuration modifies the energy spectrum of the
oscillations.  Since $P_{nlm}<T$, this process is slow in the sense
described in Section \ref{sec:RoPtSC}, and again we may use the
adiabatic invariant $E_{nlm}/\omega_{nlm}$ to estimate the  evolution
of the turbulent spectrum.  If this formation occurs at constant temperature
and mass, the mode periods are functions of $\zeta$ alone, with the
dependencies implied by Figure \ref{fig:omegas}.  Thus, the evolved
energy spectrum is given by
\begin{equation}
E_{nlm}(\zeta;\zeta_0)
=
E_{nlm}^{K/F}\frac{\omega_{nlm}(\zeta)}{\omega_{nlm}(\zeta_0)}\,,
\end{equation}
where $\zeta_0$ is the center--surface density contrast at the time that the
molecular cloud turbulence was inherited. The spectrum of supersonic
turbulence in larger scale molecular clouds is not well understood. Observations
indicate that it is close to Kolmogorov, in the range of $\sigma \sim L^{-1/3}$
to $\sigma \sim L^{-1/2}$ where $\sigma$ is the observed velocity dispersion
and $L$ the length scale. For consistency with our previous analysis,
we will consider both our Kolmogorov and Flat energy spectra as the 
inherited spectra.

For the same reason that the fundamental modes ($n=0$) and p-modes
($n>0$) produce different contributions to $\Po$, an adiabatic
contraction results in a fundamental-heavy energy spectrum if
$\zeta_0$ is sufficiently small.  This is shown at $\zeta=14$ in
Figure \ref{fig:Eevol} for both the Kolmogorov and Flat energy spectra
with $\zeta_0=1.25$.  During the collapse phase, $\Eo^{>0}/\Eo^0$ has
decreased from $3.5$ and $3.9$ to $1.3$ and $1.5$ for the
Kolmogorov and Flat energy spectra, respectively.  Thus, while in
neither case the fundamental modes dominate, in both the energy in the
fundamentals and p-modes are comparable.  This suggests that neither
$E_{nlm}^{K}$ nor $E_{nlm}^{fK}$ may be applicable, but rather some
combination of the two is appropriate for the observed pulsating
starless cores.  This also provides a natural explanation for the
dominance of low-order modes, and in particular low-order quadrupole
modes, as implicated by self-absorbed asymmetric molecular line spectra in
Barnard 68  \citep{Lada2003,Keto2006,Redman2006,Broderick2007}.

\subsection{Pulsation Damping and the Onset of Instability}
In Section \ref{sec:AME} we showed how pulsations can support
super-critical isothermal configurations.  However, in the absence
of continual excitation, these oscillations eventually decay.
The manner and timescale in which large-amplitude pulsations damp
depends upon the underlying core parameters and environment as well as
the detailed structure of each individual oscillation and their
amplitudes.  In general the lifetimes are different for each mode.
In our previous papers we have estimated life times for a few particular
cases. Figures 2 and 5 of \citet{Broderick2007} estimate a decay time
of 2 Myr for the 122 and 111 modes by mode-mode coupling.
In \citet{Broderick2008}, we estimate  decay times of 1 Myr and 0.4 Myr by 
transmission to the larger scale ISM for isolated and embedded cores
respectively. These examples are too few to define the decay times
precisely; therefore,
for simplicity, we assume that all pulsations decay
exponentially with the same lifetime, $\tau\simeq 10^6\,\yr$, chosen
to be consistent with our previous numerical simulations of the non-linear
mode-mode and core-environment couplings \citep{Broderick2007,Broderick2008}.

\begin{figure}
\begin{center}
\includegraphics[width=\columnwidth]{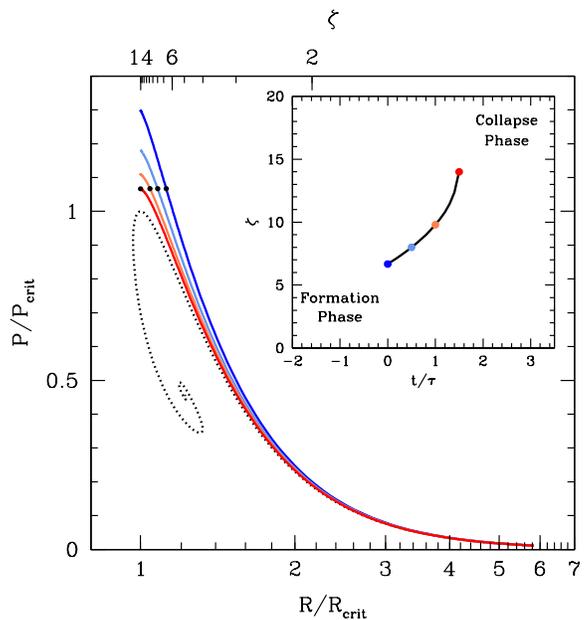}
\end{center}
\caption{Surface pressure as a function of radius for an pulsating
  isothermal gas sphere as the oscillations decay.  Initially, at
  $\zeta=1.25$, the gas has a Kolmogorov energy spectrum, evolving as
  described in Section \ref{sec:ESfRT} thereafter, and normalized such
  that at $t=0$ $\Eo=0.5\Eb$ at $\zeta=14$.  The pulsations energy decays
  exponentially in time, with decay constant $\tau$.  The dark-blue,
  light-blue, light-red and red lines show the resulting surface
  pressure at $t=0$, $0.5\tau$, $1.0\tau$ and $1.5\tau$,
  respectively.  The black points show the equilibrium configuration
  at a fixed, super-critical surface pressure.  The inset shows time
  evolution of the central density.  For reference, the points
  corresponding to the curves shown in the main figure are also
  plotted.  Finally, the relative location of the Formation and
  Collapse phases are shown.
}\label{fig:evol}
\end{figure}

The consequences of mode damping are shown explicitly in Figure
\ref{fig:evol}.  Here $P(R)$ curves are shown for a number of mode
amplitudes, where the energy spectrum is that resulting from a slow
initial contraction period, described in Section \ref{sec:ESfRT}.
Initially, these are normalized such that 50\% of the binding energy
is in oscillations at $\zeta=14$ (dark-blue line), and then at
$t/\tau=0.5$, $1.0$, and $1.5$.  Since the exterior
pressure is constant, we show a number of points with a fixed,
super-critical surface pressure, illustrating the onset of instability
during Pulsation-decay driven evolution.  The associated evolution in
$\zeta$, which in this case is simply proportional to the central
density, is shown in the inset, with the colored points corresponding
to the curves in the main Figure.

Prior to the times shown, we assume the cores were forming with
$T<\tau$, presumably during a phase of prolonged contraction
\footnote{Note that $\tau$ is also a function of the core
  density, and nonlinearly dependent upon the dynamical timescale of
  the core.  Thus, there always exist $R$ for which the mode-damping
  timescales are larger than the free-fall timescales.  This is not
  the case for the mode periods, however.}.  At some point, due
to both the increase in the dynamical importance of the pulsations
and the decrease in $\tau$ due to the increased mode amplitudes and
central density, the oscillations begin to dominate the formation
processes, and the cores enter into a pulsation-driven phase.  During
this time cores can derive a substantial portion of their pressure
support from the oscillations themselves, and their damping
drives changes in the underlying configurations on timescales
comparable to $\tau$.  Finally, for super-critical cores, once the
energy remaining in pulsations has reached some critical value,
oscillations can no longer support the core, leading to unstable
collapse.  Thus, we may think of the evolution of a pulsating
starless core as consisting of at least three phases: Formation,
Pulsation-supported and Collapse.

\subsection{Implications for the Core Mass Function}

We have already used the existence of large-amplitude oscillations to
empirically determine the relationship between the mode excitation
timescale and the other relevant timescales.  However, these
timescales are not generally independent of the core mass, and may be
violated for cores more massive than those observed, with consequences
for the core mass function.  Here we consider these within the
turbulent excitation picture.

The starless cores, or their progenitors, are constantly being
buffeted by the turbulence within the parent molecular cloud.  The
ability of a core to survive long enough to potentially form a
star  requires that the turbulence not excite 
pulsations with sufficient amplitudes to tear the core apart, either
by a single catastrophic excitation or a sequence of smaller
excitations.  We begin by noting that in the absence of oscillations there is a
well defined $M(R)$, shown in Figure \ref{fig:PvR}, and at low masses
roughly $\propto R^{-3/2}$.  When dynamically significant oscillations
are present, the core mass is a function of radius {\em and} $\Eo$.
Nevertheless,  that low mass cores are generally large and vice
versa.  As a consequence, the binding energy of cores, comparable to
$\Eb(R)\simeq GM^2/3R$ (within 30\% at all masses), is a strong
function of core mass, and in particular considerably larger for
massive, compact cores.

In contrast, the energy in turbulent motions (for Kolmogorov
turbulence) on the core scale is $\Et\propto R^{2/3}$, which decreases
with size.  This implies that for some radius, $R_0$ (or mass, $M_0$),
which depends upon the strength of the turbulence, we
have $\Eb(R_0)=\Et(R_0)$.  Since the turbulent motions fluctuate on
timescales comparable to the circulation time of a single eddy (i.e.,
$2\pi$ times the sound crossing time of the core), which is itself
comparable to the core dynamical timescale, we expect to see at least
one such excitation event for every core.  For large, low-mass cores we have
$\Eb<\Et$, and we expect the core to be torn apart when this occurs,
placing a lower-limit upon the mass of long-lived cores.
In contrast, for compact, massive cores we have $\Eb>\Et$, and
external turbulence is no longer able to excite large-amplitude
motions within the core leading again to the inherited turbulence picture
described in Section \ref{sec:ESfRT}.

The fact that $\Eb/\Et\simeq(R/R_0)^{-14/3}$ is such a strong
function of $R$ implies that the range in core radii about $R_0$, and
thus mass about $M_0$, in which large-amplitude pulsations can be
excited is narrow.  In particular,
\begin{equation}
\frac{\Delta R}{R_0}
\simeq 
\left|
\frac{d\ln R}{d\ln(\Eb/\Et)}
\right|_{R=R_0}
\simeq
0.2
\quad
\Rightarrow
\quad
\frac{\Delta M}{M_0}
\simeq
0.3\,.
\end{equation}
Cores much less massive than $M_0$ are torn apart, while those much
more massive than $M_0$ never develop the observed large-amplitude
pulsations.  The latter issue is moderated somewhat if an adiabatic
contraction phase occurs during the formation process, i.e., if the
pulsation amplitude are representative of those generated when the
core was considerably larger.  Nevertheless, the former places a hard
constraint upon the minimum mass of starless cores.

\section{Conclusions}\label{sec:C}

The onset of instability in isothermal gas spheres depends upon the
dynamical state of the object.  While this has long been known,
usually within the context of effective turbulent support, how this
occurs in practice depends upon the oscillation mode energy spectrum,
and therefore the mechanism by which pulsations are excited.  In
principle, large-amplitude breathing modes can induce instability for
masses well below $\zeta>4$  the static stability limit, a fact that is
reflective of the nature of the instability.  However, in practice,
doing so requires extraordinarily non-linear mode amplitudes, at which
point the dynamical effect is unclear.  In contrast, higher-order
pulsations are generically supportive, increasing the maximum stable
mass significantly.  For oscillation energies comparable to those
required to explain the self-absorbed, asymmetric molecular line profiles 
typically observed in starless cores, the maximum stable mass is about
 $10$--$30\%$ higher than the static case.

If the modes are excited during the core's formation, presumably
inherited from the turbulence of the parent molecular cloud, followed by an adiabatic
contraction, it is natural to expect low-$n$ oscillations to
dominate the energy spectrum.  That is, even prior to mode decay via
mode-mode or mode-environment coupling, energy from the collapse is
preferentially shunted into the long-wavelength pulsations.  Thus, it
is not unreasonable to expect starless cores to pulsate predominantly
in low-order multipole modes, with $l\ge2$.

The presence of large-amplitude oscillations can dramatically alter
the morphology of the column density maps of starless cores, mimicking the
unstable configurations of static BE spheres.  For turbulent energy spectra, dynamically
significant pulsations dramatically bias the maximum column densities
towards larger values, with the {\em average} exceeding that
associated with the critical configuration of static BE models
by factors of order unity.
This is born out by the azimuthally-averaged column density profiles,
which are very similar to isothermal gas profiles associated with
$\zeta$ above the critical value.  Despite this, the two-dimensional
column density maps are strongly asymmetric, featuring in some cases
multiple peaks, similar to those observed in practice.

The ultimate decay of the oscillations, facilitated by non-linear mode
coupling or by coupling to the surrounding molecular gas, results in
the collapse of super-critical cores.  For slow decay (lifetimes
exceeding the mode periods) this results in a sequence of quasi-stable
isothermal gas spheres, and the evolution may be determined
explicitly.  This assumes that the oscillations are not continually
re-excited by small scale processes within or near the starless cores
themselves.  However, the very existence of a significant number of
pulsating cores, with mode energies being a considerable fraction of
the cores' binding energy, argues against this.

Taken together, these suggest a  picture of
star formation,  framed entirely within a
hydrodynamical paradigm.  Self-gravitating cores condense out of the
turbulent sea within molecular clouds, born with large-amplitude
pulsations.  These pulsations drastically perturb the core shape,
mimicking non-equilibrium structures as well as initially supporting
the cores against collapse.  As the oscillations decay the core
becomes unstable resulting in star formation.  
In this scheme the rate of star formation is set not
by ambipolar diffusion of a large scale magnetic field, 
but by the non-linear decay of the core
pulsations.

We should emphasize that while we have framed this new picture
entirely in terms of hydrodynamics, small scale tangled magnetic fields 
in equipartition with the turbulent energy are
still allowed and might be required.  
The model of oscillations as perturbations does  not allow amplitudes
large enough to produce the supersonic motions occasionally
seen in a few cores. These are the exception rather than the rule
and might not be pulsations at all, but rather inward motions
as the 
core center transitions to free-fall collapse.
 However, if
these motions are sub-Alfv\'enic pulsations, they may be subsumed into an
equivalent magnetohydrodynamic formulation, in which we consider only
small-scale magnetic fields.  In this case, it would be the decay of the
magnetohydrodynamic pulsations, not ambipolar diffusion, that sets the star
formation rate.  Nevertheless, as empirical evidence accumulates for
the existence of large-amplitude oscillations in starless cores, it is
clear that their dynamical effects upon the cores themselves cannot be
ignored.


\begin{thebibliography}{99}
\expandafter\ifx\csname natexlab\endcsname\relax\def\natexlab#1{#1}\fi

\bibitem[{{Aguti} {et~al.}(2007)}]{Aguti2007}
{Aguti}, E.~D. {et~al.} 2007, \apj, 665, 457

\bibitem[{{Alves} {et~al.}(2001){Alves}, {Lada}, \& {Lada}}]{AlvesLadaLada2001}
{Alves}, J.~F., {Lada}, C.~J., \& {Lada}, E.~A. 2001, \nat, 409, 159

\bibitem[{{Andre} {et~al.}(1996){Andre}, {Ward-Thompson}, \&
  {Motte}}]{Andre1996}
{Andre}, P., {Ward-Thompson}, D., \& {Motte}, F. 1996, \aap, 314, 625

\bibitem[{{Bacmann} {et~al.}(2000)}]{Bacmann2000}
{Bacmann}, A. {et~al.} 2000, \aap, 361, 555

\bibitem[{{Beichman} {et~al.}(1986)}]{Beichman1986}
{Beichman}, C.~A. {et~al.} 1986, \apj, 307, 337

\bibitem[{{Benson} \& {Myers}(1989)}]{BensonMyers1989}
{Benson}, P.~J. \& {Myers}, P.~C. 1989, \apjs, 71, 89

\bibitem[{{Bergin} \& {Tafalla}(2007)}]{BerginTafalla2007}
{Bergin}, E.~A. \& {Tafalla}, M. 2007, \araa, 45, 339

\bibitem[{{Bergin} {et~al.}(2006)}]{Bergin2006}
{Bergin}, E.~A. {et~al.} 2006, \apj, 645, 369

\bibitem[{{Bonnor}(1956)}]{Bonnor1956}
{Bonnor}, W.~B. 1956, \mnras, 116, 351

\bibitem[{{Broderick} {et~al.}(2007){Broderick}, {Keto}, {Lada}, \&
  {Narayan}}]{Broderick2007}
{Broderick}, A.~E., {Keto}, E., {Lada}, C.~J., \& {Narayan}, R. 2007, \apj,
  671, 1832

\bibitem[{{Broderick} {et~al.}(2008){Broderick}, {Narayan}, {Keto}, \&
  {Lada}}]{Broderick2008}
{Broderick}, A.~E., {Narayan}, R., {Keto}, E., \& {Lada}, C.~J. 2008, \apj,
  682, 1095

\bibitem[{{Crapsi} {et~al.}(2007){Crapsi}, {Caselli}, {Walmsley}, \&
  {Tafalla}}]{Crapsi2007}
{Crapsi}, A., {Caselli}, P., {Walmsley}, M.~C., \& {Tafalla}, M. 2007, \aap,
  470, 221

\bibitem[{{di Francesco} {et~al.}(2007)}]{DiFrancesco2007}
{di Francesco}, J. {et~al.} 2007, in Protostars and Planets V, ed.
  B.~{Reipurth}, D.~{Jewitt}, \& K.~K. (University of Arizona Press: Tuscon),
  17--32

\bibitem[{{Dickman} \& {Clemens}(1983)}]{DickmanClemens1983}
{Dickman}, R.~L. \& {Clemens}, D.~P. 1983, \apj, 271, 143

\bibitem[{{Evans} {et~al.}(2001){Evans}, {Rawlings}, {Shirley}, \&
  {Mundy}}]{Evans2001}
{Evans}, II, N.~J., {Rawlings}, J.~M.~C., {Shirley}, Y.~L., \& {Mundy}, L.~G.
  2001, \apj, 557, 193

\bibitem[{{Field} {et~al.}(2008){Field}, {Blackman}, \& {Keto}}]{Field2008}
{Field}, G.~B., {Blackman}, E.~G., \& {Keto}, E.~R. 2008, \mnras, 385, 181

\bibitem[{{Gingold} \& {Monaghan}(1980)}]{Ging-Mona:80}
{Gingold}, R.~A. \& {Monaghan}, J.~J. 1980, \mnras, 191, 897

\bibitem[{{Goldstein}(1980)}]{Goldstein1980}
{Goldstein}, H. 1980, {Classical Mechanics, 2nd ed.} (Addison-Wesley: Reading,
  Mass.)

\bibitem[{{Gon{\c c}alves} {et~al.}(2004){Gon{\c c}alves}, {Galli}, \&
  {Walmsley}}]{Goncalves2004}
{Gon{\c c}alves}, J., {Galli}, D., \& {Walmsley}, M. 2004, \aap, 415, 617

\bibitem[{{Jessop} \& {Ward-Thompson}(2000)}]{JessopWT2000}
{Jessop}, N.~E. \& {Ward-Thompson}, D. 2000, \mnras, 311, 63

\bibitem[{{Kandori} {et~al.}(2005)}]{Kandori2005}
{Kandori}, R. {et~al.} 2005, \aj, 130, 2166

\bibitem[{{Keto} {et~al.}(2006){Keto}, {Broderick}, {Lada}, \&
  {Narayan}}]{Keto2006}
{Keto}, E., {Broderick}, A.~E., {Lada}, C.~J., \& {Narayan}, R. 2006, \apj,
  652, 1366

\bibitem[{{Keto} \& {Caselli}(2008)}]{KetoCaselli2008}
{Keto}, E. \& {Caselli}, P. 2008, \apj, 683, 238

\bibitem[{{Keto} \& {Caselli}(2010)}]{KetoCaselli2010}
{Keto}, E. \& {Caselli}, P. 2010, \mnras, 402, 1625

\bibitem[{{Keto} \& {Field}(2005)}]{KetoField2005}
{Keto}, E. \& {Field}, G. 2005, \apj, 635, 1151

\bibitem[{{Kirk} {et~al.}(2005){Kirk}, {Ward-Thompson}, \&
  {Andr{\'e}}}]{KirkWTAndre2005}
{Kirk}, J.~M., {Ward-Thompson}, D., \& {Andr{\'e}}, P. 2005, \mnras, 360, 1506

\bibitem[{{Lada} {et~al.}(2003){Lada}, {Bergin}, {Alves}, \&
  {Huard}}]{Lada2003}
{Lada}, C.~J., {Bergin}, E.~A., {Alves}, J.~F., \& {Huard}, T.~L. 2003, \apj,
  586, 286

\bibitem[{{Lada} {et~al.}(2008)}]{Lada2008}
{Lada}, C.~J. {et~al.} 2008, \apj, 672, 410

\bibitem[{{Landau} \& {Lifshitz}(1976)}]{LandauLifshitz1976}
{Landau}, L.~D. \& {Lifshitz}, E.~M. 1976, {Mechanics, 3rd ed.}, ed. {Landau,
  L.~D.~\& Lifshitz, E.~M.} (Butterworth-Heinemann Ltd: Oxford)

\bibitem[{{Myers} \& {Benson}(1983)}]{MyersBenson1983}
{Myers}, P.~C. \& {Benson}, P.~J. 1983, \apj, 266, 309

\bibitem[{{Myers} {et~al.}(1983){Myers}, {Linke}, \&
  {Benson}}]{MyersLinkeBenson1983}
{Myers}, P.~C., {Linke}, R.~A., \& {Benson}, P.~J. 1983, \apj, 264, 517

\bibitem[{{Pagani} {et~al.}(2003)}]{Pagani2003}
{Pagani}, L. {et~al.} 2003, \aap, 406, L59

\bibitem[{{Pagani} {et~al.}(2004)}]{Pagani2004}
{Pagani}, L. {et~al.} 2004, \aap, 417, 605

\bibitem[{{Pineda} {et~al.}(2010)}]{Pineda2010}
{Pineda}, J.~E. {et~al.} 2010, \apjl, 712, L116

\bibitem[{{Rathore} {et~al.}(2003){Rathore}, {Broderick}, \&
  {Blandford}}]{Rath-Brod-Blan:03}
{Rathore}, Y., {Broderick}, A.~E., \& {Blandford}, R. 2003, \mnras, 339, 25

\bibitem[{{Redman} {et~al.}(2006){Redman}, {Keto}, \& {Rawlings}}]{Redman2006}
{Redman}, M.~P., {Keto}, E., \& {Rawlings}, J.~M.~C. 2006, \mnras, 370, L1

\bibitem[{{Schnee} {et~al.}(2007)}]{Schnee2007}
{Schnee}, S. {et~al.} 2007, \apj, 671, 1839

\bibitem[{{Shirley} {et~al.}(2002){Shirley}, {Evans}, \&
  {Rawlings}}]{ShirleyEvansRawlings2002}
{Shirley}, Y.~L., {Evans}, II, N.~J., \& {Rawlings}, J.~M.~C. 2002, \apj, 575,
  337

\bibitem[{{Sohn} {et~al.}(2007)}]{Sohn2007}
{Sohn}, J. {et~al.} 2007, \apj, 664, 928

\bibitem[{{Stamatellos} \& {Whitworth}(2003)}]{StamatellosWhitworth2003}
{Stamatellos}, D. \& {Whitworth}, A.~P. 2003, \aap, 407, 941

\bibitem[{{Tafalla} {et~al.}(2002){Tafalla}, {Myers}, {Caselli}, {Walmsley}, \&
  {Comito}}]{Tafalla2002}
{Tafalla}, M., {Myers}, P.~C., {Caselli}, P., {Walmsley}, C.~M., \& {Comito},
  C. 2002, \apj, 569, 815

\bibitem[{{Teixeira} {et~al.}(2005){Teixeira}, {Lada}, \&
  {Alves}}]{Teixeira2005}
{Teixeira}, P.~S., {Lada}, C.~J., \& {Alves}, J.~F. 2005, \apj, 629, 276

\bibitem[{{Ward-Thompson} {et~al.}(2002){Ward-Thompson}, {Andr{\'e}}, \&
  {Kirk}}]{Ward-Thompson2002}
{Ward-Thompson}, D., {Andr{\'e}}, P., \& {Kirk}, J.~M. 2002, \mnras, 329, 257

\bibitem[{{Ward-Thompson} {et~al.}(1999){Ward-Thompson}, {Motte}, \&
  {Andre}}]{Ward-Thompson1999}
{Ward-Thompson}, D., {Motte}, F., \& {Andre}, P. 1999, \mnras, 305, 143

\bibitem[{{Ward-Thompson} {et~al.}(1994){Ward-Thompson}, {Scott}, {Hills}, \&
  {Andre}}]{Ward-Thompson1994}
{Ward-Thompson}, D., {Scott}, P.~F., {Hills}, R.~E., \& {Andre}, P. 1994,
  \mnras, 268, 276

\bibitem[{{Ward-Thompson} {et~al.}(2007)}]{Ward-Thompson2007}
{Ward-Thompson}, D. {et~al.} 2007, in Protostars and Planets V, ed.
  B.~{Reipurth}, D.~{Jewitt}, \& K.~K. (University of Arizona Press: Tuscon),
  33--46

\bibitem[{{Zucconi} {et~al.}(2001){Zucconi}, {Walmsley}, \&
  {Galli}}]{Zucconi2001}
{Zucconi}, A., {Walmsley}, C.~M., \& {Galli}, D. 2001, \aap, 376, 650

\end{thebibliography}
\end{document}